\documentclass[aps,prc,preprint,draft,showpacs,showkeys,tightenlines]{revtex4}

\usepackage{epsf} 
 
\begin{document}  
 
\title{ Assessment of uncertainties in QRPA $0\nu\beta\beta$-decay 
    nuclear matrix elements}

\date{\today}  
 
\author{V. A. Rodin}  
\email{vadim.rodin@uni-tuebingen.de} 
\affiliation{Institute f\"{u}r Theoretische Physik der Universit\"{a}t 
T\"{u}bingen, D-72076 T\"{u}bingen, Germany} 
\author{Amand Faessler} 
\email{amand.faessler@uni-tuebingen.de} 
\affiliation{Institute f\"{u}r Theoretische Physik der Universit\"{a}t 
T\"{u}bingen, D-72076 T\"{u}bingen, Germany} 
\author{F. \v Simkovic} 
\email{fedor.simkovic@fmph.uniba.sk} 
\altaffiliation{On  leave of absence from Department of Nuclear 
Physics, Comenius University, Mlynsk\'a dolina F1, SK--842 15 
Bratislava, Slovakia}  
\affiliation{Institute f\"{u}r Theoretische Physik der Universit\"{a}t 
T\"{u}bingen, D-72076 T\"{u}bingen, Germany} 
\author{Petr Vogel} 
\email{pxv@caltech.edu} 
\affiliation{Kellogg Radiation Laboratory 106-38, California Institute 
of Technology, Pasadena, CA 91125, USA and \\ 
Physics Department, Stanford University, Stanford, CA 94305, USA} 
 
\begin{abstract} 
The nuclear matrix elements $M^{0\nu}$  of the 
neutrinoless double beta decay ($0\nu\beta\beta$)  
of most nuclei with known $2\nu\beta\beta$-decay rates are systematically evaluated 
using the Quasiparticle Random Phase Approximation 
(QRPA) and Renormalized QRPA (RQRPA).  
The experimental $2\nu\beta\beta$-decay rate is used  
to adjust the most relevant parameter, the strength of  
the particle-particle interaction.  
New results confirm that with such procedure the  
$M^{0\nu}$ values become essentially independent on the size of the  
single-particle basis. 
Furthermore, the matrix elements are shown to be also rather stable with respect  
to the possible quenching of the axial  
vector strength parametrized by reducing the 
coupling constant $g_A$,  as well as to the uncertainties of  
parameters describing the short range nucleon correlations.  
Theoretical arguments in favor of the adopted way of determining 
the interaction parameters are presented. 
Furthermore, a discussion of other implicit and explicit parameters,
inherent to the QRPA method, is presented. Comparison is made of the ways
these factors are chosen by different authors. It is suggested that most of
the spread among the published $0\nu\beta\beta$ decay nuclear matrix elements
can be ascribed to these choices.
\end{abstract}  
 
 

\maketitle 
 
 
\section{Introduction} 
 
Inspired by the spectacular discoveries of oscillations of atmospheric 
\cite{fukuda98}, solar \cite{clev98,abdu99,fukuda02,ahmad01} , 
and reactor neutrinos\cite{eguchi03} (for recent reviews see  
\cite{McKeown04,Concha03,Grimus03,bil03,Langacker04})  
the physics community worldwide is embarking on 
the next challenging problem, finding whether neutrinos are indeed  
Majorana particles as many particle physics models suggest.  
Study of the neutrinoless 
double beta decay ($0\nu\beta\beta$) is the best potential source 
of information about the Majorana nature of the neutrinos 
\cite{FS98,V02,EV2002,EE04}. Moreover, 
the rate of the  $0\nu\beta\beta$ decay, or limits on its value, can 
be used to constrain the neutrino mass pattern and the absolute 
neutrino mass scale, i.e., information not available by the study of 
neutrino oscillations. (The goals, and possible future directions 
of the field are described, e.g., in the recent study \cite{matrix}. The 
issues particularly relevant for the program of $0\nu\beta\beta$ decay 
search are discussed in \cite{nustudy}.)     
 
The observation of $0\nu\beta\beta$ decay would immediately tell us that 
neutrinos are massive Majorana particles. But without accurate calculations 
of the nuclear matrix elements it will be difficult to reach quantitative 
conclusions about the absolute neutrino masses and mass hierarchies 
and confidently plan new experiments.  
Despite years of effort there is at present a lack of consensus  
among nuclear theorists how to correctly calculate the nuclear matrix 
elements, and how to estimate their uncertainty (see e.g. \cite{EE04,bah04}). 
Since an overwhelming majority of published calculations is based 
on the Quasiparticle Random Phase Approximation (QRPA) and its modifications, 
it is worthwhile to try to see what causes the sizable spread of the 
calculated $M^{0\nu}$ values. Does it reflect some fundamental uncertainty, 
or is it mostly related to different choices of various adjustable 
parameters? If the latter is true (and we believe it is) can one 
find and justify an optimal choice that largely removes such  
unphysical dependence? 
 
In the previous paper \cite{Rod03a} we have shown that by adjusting the 
most important parameter, the strength of the isoscalar particle-particle 
force so that the known rate of the $2\nu\beta\beta$-decay is correctly 
reproduced, the dependence of the calculated $0\nu\beta\beta$ 
nuclear matrix elements $M^{0\nu}$ on other things that are not a priori 
fixed, is essentially removed. In particular, we have shown that this  
is so as far the number of single particle states included is concerned, 
and the choice of the different realistic representations of the 
nucleon $G$-matrix. In \cite{Rod03a} we applied this procedure 
to the  $0\nu\beta\beta$ decay candidate nuclei, 
$^{76}$Ge, $^{100}$Mo, $^{130}$Te, and $^{136}$Xe. 
 
In the present work we wish to expand and better justify the ideas 
presented in \cite{Rod03a}. First, the method is systematically applied  
to calculate the nuclear matrix elements $M^{0\nu}$ 
for most of the nuclei with known experimental $2\nu\beta\beta$-decay rates.  
Second, the sensitivity of the results to variation of other model parameters   
is tested. These are the axial vector  
quenching factor, commonly described as a modification of 
the constant  $g_A$, and the parameters that describe the effect of the short  
range correlations. Finally, arguments in favor of the chosen  
calculation method are presented and discussed.

\section{Details of the calculation of $0\nu\beta\beta$ decay matrix elements} 
 
Provided that a virtual light Majorana neutrino with  the effective mass  
$\langle m_{\beta\beta} \rangle$, 
\begin{equation} 
\langle m_{\beta\beta} \rangle = \sum_i^N |U_{ei}|^2 e^{i\alpha_i} m_i ~, 
~({\rm all~} m_i \ge 0)~, 
\end{equation} 
is exchanged between the nucleons \cite{foot1} 
the half-life of the $0\nu\beta\beta$ decay is given by  
\begin{equation} 
\frac{1}{T_{1/2}} = G^{0\nu}(E_0,Z) |{M'}^{0\nu}|^2  
|\langle m_{\beta\beta} \rangle|^2~, 
\end{equation} 
where $G^{0\nu}(E_0,Z)$ is the precisely calculable phase-space factor, 
and ${M'}^{0\nu}$ is the corresponding nuclear matrix element.  
Thus, obviously, any uncertainty in ${M'}^{0\nu}$ makes the value 
of $\langle m_{\beta\beta} \rangle$ equally uncertain. 
 
The elements of the 
mixing matrix $|U_{ei}|^2$ and the mass-squared differences $\Delta m^2$ 
can be determined in oscillation experiments.  
If the existence of the $0\nu\beta\beta$ decay is proved and the  
value of $T_{1/2}$ is found, combining the  knowledge of the first 
row of the neutrino 
mixing matrix $|U_{ei}|^2$ and the mass-squared differences $\Delta m^2$, 
a relatively narrow range of 
absolute neutrino mass scale can be determined, independently of the 
Majorana phases $\alpha_i$ in most situations \cite{EV2002,PP2002}. However, 
such an important insight is possible only if the nuclear matrix elements 
are accurately known.  
 
The nuclear matrix element ${M'}^{0\nu}$ is defined as 
\begin{equation} 
{M'}^{0\nu} = \left(\frac{g_A}{1.25}\right)^2~\langle f  
| -\frac{M^{0\nu}_F}{g^2_A} + M^{0\nu}_{GT} +  
M^{0\nu}_T |i\rangle  
\label{eq:m0nudef} 
\end{equation} 
where $|i\rangle, ~(|f\rangle )$ are the wave functions of the ground 
states of the initial (final) nuclei.  We note that for $g_A=1.25$  
nuclear matrix element ${M'}^{0\nu}$ coincides with ${M}^{0\nu}$ 
of our previous work \cite{Rod03a}. We have chosen this, somewhat awkward,  
parameterization so that we could later modify the value of $g_A$ and still 
use the same phase space factor $G^{0\nu}(E_0,Z)$  
that contains $g_A^4 = (1.25)^4$, 
tabulated e.g. in Ref. \cite{si99}. 
 
The explicit forms of the operators $M^{0\nu}_F, M^{0\nu}_{GT}$ and 
$M^{0\nu}_T$ are given in Ref.~\cite{si99}.  
In order to explain the notation used below we summarize here only the most 
relevant formulae. We begin with the effective transition operator  
in the momentum representation 
\begin{equation} 
\Omega  =  \tau^+ \tau^+  
\frac{(-h_F + h_{GT}\sigma_{12} - h_T S_{12})} 
{q (q+E^m_J-(E^i_{g.s.}+E^f_{g.s})/2)} 
 ~,~~ 
\sigma_{12} = \vec{\sigma_1} \cdot \vec{\sigma_2} ~,~~ 
S_{12} = 3\vec{\sigma_1} \cdot \vec{q} \vec{\sigma_2} \cdot \vec{q} 
-\sigma_{12} ~.  
\end{equation} 
Here, $h_F(q^2) = g_V^2(q^2),$ and  
\begin{equation} 
h_{GT}  =   g_A^2 \left[ 1 - \frac{2}{3} \frac{\vec{q}^2} 
{\vec{q}^2 + m_{\pi}^2} + \frac{1}{3} \left( \frac{\vec{q}^2} 
{\vec{q}^2 + m_{\pi}^2} \right)^2 \right],  
h_T  =   g_A^2 \left[\frac{2}{3} \frac{\vec{q}^2} {\vec{q}^2 + m_{\pi}^2} 
- \frac{1}{3} \left( \frac{\vec{q}^2}{\vec{q}^2 + m_{\pi}^2}  
\right)^2 \right]. 
\label{eq:formf} 
\end{equation}  
For simplicity we do not explicitly indicate here the $q^2$ dependence of the 
form factor $g_A(q^2)$ (the usual dipole form is used for both 
$g_V(q^2)$ and $g_A(q^2)$), and the terms containing $1/m_p^2$. The 
full expressions can be found in Ref.~\cite{si99} and are used in 
the numerical calculations. Note that the space part of the  
momentum transfer four-vector  
is used in Eq. (\ref{eq:formf}) since the time component is much smaller than 
$q \equiv |\vec{q}|$. 
 
The parts containing $\vec{q}^2 + m_{\pi}^2$ come from 
the induced pseudoscalar form factor $g_P$ for which the partially 
conserved axial-vector current hypothesis (PCAC) has been used.  
In comparison with most of previous $0\nu\beta\beta$ decay studies  
\cite{SSPF1997,VZ86,CFT87,MK88} the higher  
order terms of the nucleon current (in particular the induced 
pseudoscalar as well as the weak magnetism) are  included 
in the present calculation, resulting  
in a reduction of the nuclear matrix element by about 30\% \cite{si99}. 
In the numerical calculation of ${M'}^{0\nu}$ here the summation  
over the virtual states in the intermediate nucleus is explicitly 
performed. 
 
The relatively important role of the induced nucleon currents deserves 
a comment. In the charged current neutrino induced reactions the 
parts of the cross section containing $g_P$ are usually unimportant 
because they appear proportional to the outgoing charged lepton mass. 
That comes about because the equation of motion for the  
outgoing on-mass-shell 
lepton is used. Here, this cannot be done since the neutrino is virtual, 
and highly off-mass-shell. 
 
In order to calculate the matrix element in the coordinate space, one 
has to evaluate first the `neutrino potentials' that in our case 
explicitly depend on the energy $E^m$ of the intermediate state, 
\begin{equation} 
H_K (r_{12}) = \frac{2}{\pi g_A^2} {R} \int_0^{\infty}~ f_K(qr_{12})~  
\frac{ h_K (q^2) q dq } 
{q + E^m - (E_i + E_f)/2} ~, 
\end{equation} 
where $K$ stands for $F, GT, T$.  $f_{F,GT}(qr_{12}) = j_0(qr_{12})$ and  
$f_{T}(qr_{12}) = - j_2(qr_{12})$, where $j_{0,2}$  are the spherical Bessel  
functions. The form factor combinations $h_K$ 
are defined in~ Eq.(\ref{eq:formf}), $r_{12}$ is the distance between the  
nucleons, $R$ is the nuclear radius,  
and $E_i (E_f)$ are the ground state energies of the 
initial (final) nuclei. We note that in Ref. \cite{si99}  
the tensor potential is presented incorrectly with the spherical Bessel function  
$j_0(x)=\sin(x)/x$ instead of $j_2(x)$. However, the numerical results were  
obtained with the correct expression.   
 
The individual parts of the matrix element ${M'}^{0\nu}$, Eq.(\ref{eq:m0nudef}), 
are given by the expression: 
\begin{eqnarray} 
M_K = & & \sum_{J^{\pi},k_i,k_f,\mathcal{J}} \sum_{pnp'n'} 
(-1)^{j_n + j_{p'} + J + {\mathcal J}} \sqrt{ 2 {\mathcal J} + 1} 
\left\{ 
\begin{array}{c c c} 
j_p & j_n & J  \\ 
 j_{n'} & j_{p'} & {\mathcal J}  
\end{array} 
\right\}  \nonumber \\ 
& & 
\langle p(1), p'(2); {\mathcal J} \parallel f(r_{12})  
O_K f(r_{12}) \parallel n(1), n'(2); {\mathcal J} \rangle \nonumber \\ 
& & \times \langle 0_f^+ ||  
[ \widetilde{c_{p'}^+ \tilde{c}_{n'}}]_J || J^{\pi} k_f \rangle 
\langle  J^{\pi} k_f |  J^{\pi} k_i \rangle 
 \langle  J^{\pi} k_f || [c_p^+ \tilde{c}_n]_J || 0_i^+ \rangle ~.  
\label{eq:long} 
\end{eqnarray} 
Here we define only those symbols that are needed further, for full 
explanation see \cite{si99}. The summation over $J^{\pi}$ represents the 
summation over the different multipolarities in the virtual intermediate 
odd-odd nucleus; such states are labeled by the indices $k_i,k_f$ 
when they are reached by the corresponding one-body operators acting 
on the initial or final nucleus, respectively. The overlap factor 
$\langle  J^{\pi} k_f |  J^{\pi} k_i \rangle$  accounts for the 
difference between them. 
 
The operators $O_K, K = F,GT,T$ contain the corresponding neutrino 
potentials and the relevant spin and isospin operators. Short range 
correlation of the two initial neutrons and two final protons 
are described by the Jastrow-like function $f(r_{12})$, 
\begin{equation} 
f(r_{12}) = 1 - e^{-\gamma_1 r_{12}^2}(1 - \gamma_2 r_{12}^2) ~, 
\label{eq:short} 
\end{equation} 
where the usual choice is \cite{MS1976} $\gamma_1$ = 1.1 fm$^2$, 
$\gamma_2$ = 0.68 fm$^2$. (These two parameters are correlated.) 
 
Finally, the reduced matrix elements of the one-body operators 
$c_p^+ \tilde{c}_n$ ($\tilde{c}_n$ denotes the time-reversed state) 
depend on the BCS coefficients $u_i,v_j$ and on the QRPA vectors 
$X,Y$ \cite{si99}. The difference between QRPA and RQRPA resides 
in the way these reduced matrix elements are calculated. 
 
As in~\cite{Rod03a}, the quasiparticle random phase approximation (QRPA) 
and its modification, the renormalized QRPA (RQRPA), are used to describe  
the structure of the intermediate nuclear states virtually excited in the  
double beta decay. We stress  
that in the QRPA and RQRPA one can include essentially unlimited set  
of single-particle states, labeled by $p$, $n$ in Eq.(\ref{eq:long}),  
but only a limited subset of configurations  
(iterations of the particle-hole, 
respectively two-quasiparticle configurations),  
in contrast to the nuclear shell model where the opposite is true. 
On the other hand, within the QRPA there is no obvious procedure  
that determines how many single particle states one should include. 
Hence, various authors choose this crucial number ad hoc, basically 
for reasons of convenience.

As has been already shown in~\cite{Rod03a}, a particular choice  
of the realistic  
residual two-body interaction potential  has almost no impact on  
the finally calculated mean value and variance $\sigma$ 
of ${M'}^{0\nu}$, with the overwhelming contribution to $\sigma$ coming 
from the choice of the single-particle basis size. Therefore, we perform the  
calculations here using $G$-matrix based only on the Bonn-CD  
nucleon-nucleon potential. 
 
It is well known that the residual interaction is an effective interaction  
depending on the size of the 
single-particle (s.p.) basis. Hence, when the basis is changed, the interaction 
should be modified as well.  
For each nucleus in question three single-particle bases are chosen  
as described in Ref. \cite{Rod03a} 
with the smallest set corresponding to $1 \hbar\omega$ particle-hole excitations,  
and the largest   
to about $4 \hbar\omega$ excitations. The s.p. energies are calculated 
with the Coulomb corrected Woods-Saxon potential. 
 
\subsection{Parameter adjustment} 
 
In QRPA and RQRPA there are three important global parameters  
renormalizing the bare residual interaction. 
First, the pairing part of the interaction is multiplied by a factor $g_{pair}$ 
whose magnitude is adjusted, for both protons and neutrons 
separately, such that the pairing gaps for the initial and final nuclei  
are correctly reproduced. 
This is a standard procedure and  
it is well-known that within the BCS method the strength of  
the pairing interaction depends on the size 
of the s.p. basis. 
 
Second, the particle-hole interaction block is renormalized by an overall strength 
parameter $g_{ph}$ which is typically adjusted by requiring 
that the energy of the giant GT resonance is correctly reproduced.  
We find that the calculated energy of the giant GT state is almost independent  
of the size of the s.p. basis and is well reproduced with $g_{ph} \approx 1$.  
Accordingly, we use $g_{ph} = 1$ throughout, without adjustment. 
 
Third, an important strength parameter $g_{pp}$ renormalizes  
the particle-particle interaction 
(the importance of the particle-particle interaction 
for the $\beta$ strength was recognized first in Ref. \cite{Cha83}, and 
for the $\beta\beta$ decay in \cite{VZ86}).  
The decay rate for both modes of $\beta\beta$ decay is well known to depend sensitively 
on the value of $g_{pp}$. 
This property has been used in~\cite{Rod03a} to  
fix the value of $g_{pp}$ for each of the s.p. bases 
so that the known half-lives of the $2\nu \beta\beta$ decay are correctly reproduced.  
 
Such an adjustment of $g_{pp}$, 
when applied to all multipoles $J^{\pi}$, has been shown in~\cite{Rod03a} 
to remove much of the sensitivity to the number of single-particle states,  
to the $NN$ potential employed, and even to whether RQRPA or just simple 
QRPA methods are used. This is in contrast to typical conclusion made in the recent past 
\cite{SSPF1997,SK01,CS03} that the values of ${M'}^{0\nu}$ vary substantially depending  
on all of these things. 
 
We believe that the $2\nu$ decay rate is especially suitable for such an adjustment, 
in particular because it involves the same initial and final states as the 
$0\nu$ decay. Moreover, the QRPA is a method designed to describe collective states as 
well as to obey various sum rules. Both double-beta decay amplitudes, 
$0\nu\beta\beta$ and $2\nu\beta\beta$, receive contributions from many  
intermediate states 
and using one of them for fixing parameters of QRPA seems preferable. 
We will elaborate this point in the next section.   
 
While the above arguments are plausible, the issue of adequacy of the QRPA 
method in general, and the chosen way of adjusting the parameters, should 
be also tested using suitable simplified models that allow exact solution. 
One such test was performed recently \cite{model}. It involved two 
shells of varying separation, and a schematic interaction. From the 
point of view of the present work, the model suggested that QRPA is an 
excellent approximation of the exact solution. However, it turned out that 
the present method of parameter adjustment was not able to eliminate fully the 
effect of the higher, almost empty,  shell on the $0\nu\beta\beta$ decay 
matrix element. It is not clear whether this is a consequence 
of the schematic nature of the model, or of some more fundamental cause. 
In any case, the present results suggest that, within QRPA and RQRPA, 
the effects of the far away single particle states can be indeed eliminated, 
or at least substantially reduced.  
 
It is well known that the calculated Gamow-Teller strength is larger 
than the experimental one. To account for this, it is customary to `quench' 
the calculated GT matrix elements. Formally, this could be conveniently accomplished 
by replacing the true value of the coupling constant $g_A$ = 1.25 by a quenched 
value $g_A \simeq$ 1.0. It is not clear whether similar phenomenon exists 
for other multipoles, besides $J^{\pi} = 1^+$. To see the dependence on 
the chosen $g_A$ value, we use  
in this work both the unquenched and quenched value of  
the axial current coupling constant  
$g_A=1.25$ and $g_A=1.0$, respectively (for all multipoles). 
The matrix elements ${M'}^{0\nu}$ calculated for the three 
s.p. bases and a 
fixed $g_A$ are relatively close to each other. 
As in~\cite{Rod03a}, for each nucleus the corresponding 
average  $\langle {M'}^{0\nu} \rangle$ matrix elements  
(averaged over the three choices of the s.p. space)  
is evaluated, as well as its variance $\sigma$. 
These quantities (with the value of $\sigma$ in parentheses)  
are shown in Table \ref{tab:t12}, columns 4 and 5. 
Two lines for each nucleus represent the results obtained  
with $g_A=1.25$ (the upper one) and  
$g_A=1.0$ (the lower one). One can see that not only is the variance  
substantially less than the average value, but the results of QRPA,  
albeit slightly larger, are quite 
close to the RQRPA values.  
Furthermore, the ratio of the  
matrix elements calculated with different $g_A$ 
is closer to unity (in most cases they differ only by $\sim$20\%)  
than the ratio of the respective $g_A$ squared (1.6 in our case).  
The reason for such a partial compensation of the $g_A$-dependence is that  
the experimental $M^{2\nu}$ for $g_A=1.0$ is larger than $M^{2\nu}$  
for $g_A=1.25$ after adjusting in both calculation the $g_{pp}$  
separately to the experimental $2\nu\beta\beta$-decay transition probability. 
Correspondingly, one gets smaller adjusted value  
of $g_{pp}$ leading to larger calculated $M^{0\nu}$.  
Thus, with the adopted choice of parameter fixing the resulting $0\nu\beta\beta$ 
decay rate depends on the adopted $g_A$ markedly less than the naive scaling  
$g_A^4$ that would suggest a change by a factor of 2.44. 
 
Naturally, the $2\nu\beta\beta$ half-lives are known only with some uncertainty. 
To see how the experimental error in $M^{2\nu}$ affects the calculated ${M'}^{0\nu}$, 
the derivatives $\frac{d{M'}^{0\nu}}{dM^{2\nu}}$ at the experimental value of $M^{2\nu}$  
are calculated. Finally, the errors induced by the experimental uncertainties  
in  $M^{2\nu}$, $\varepsilon_{exp.}=\frac{d{M'}^{0\nu}}{dM^{2\nu}}\delta M^{2\nu}_{exp}$,  
are given in column 6 of Table~\ref{tab:t12}.

\begin{figure}[t] 
\begin{center} 
    \leavevmode 
    \epsfxsize=.9\textwidth 
    \epsffile{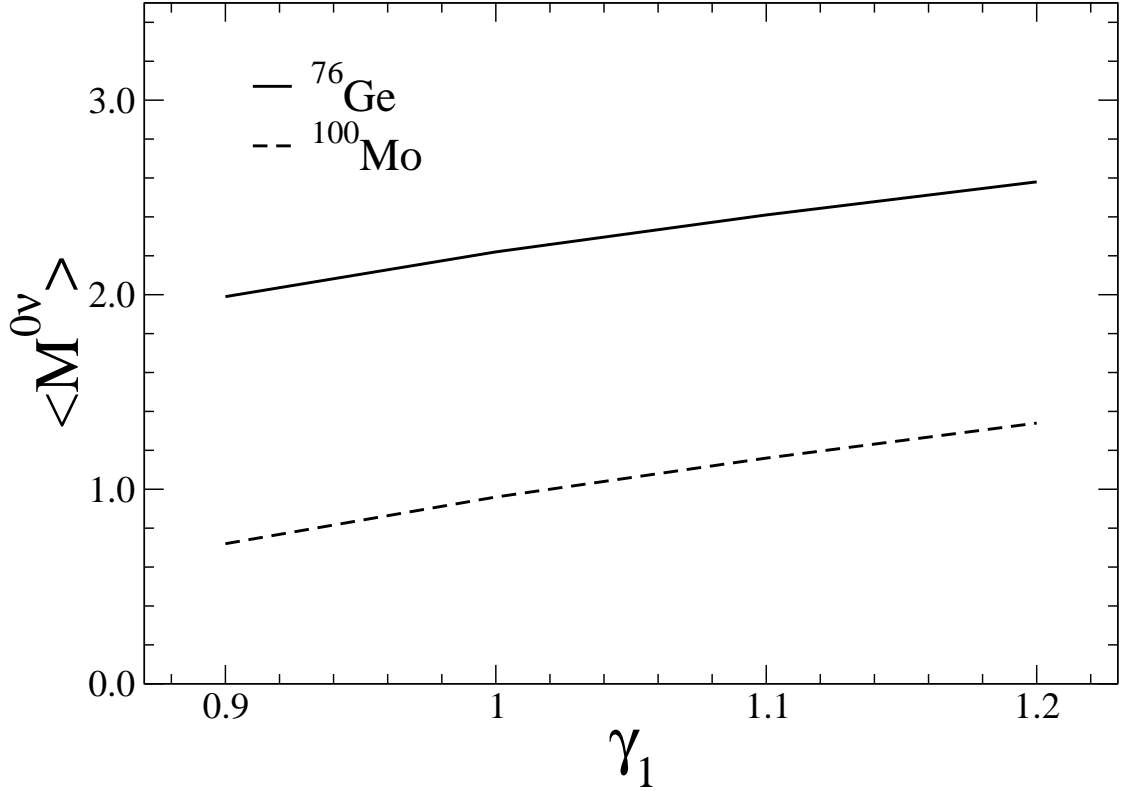} 
\caption{Nuclear matrix element 
${M'}^{0\nu}$ as function of parameter $\gamma_1$ of the function $f(r_{12})$, 
Eq. (\ref{eq:short}),  
used to account for the two-nucleon short range correlations.
Calculation performed in RQRPA with $g_{pp}= 0.99~(^{76}Ge),~1.21~(^{100}Mo)$
 and the small single particle level scheme. }
   \label{0nbbfig:1} 
\end{center} 
\end{figure}

Another uncertainty is related to the treatment of the short range nucleon 
correlation. The adopted form of the Jastrow-type factor $f(r_{12})$,  
Eq. (\ref{eq:short}), 
is based on the work \cite{MS1976} where a range of the values of 
the parameter $\gamma_1$ is given. 
We vary that parameter $\gamma_1$ from 0.9 to 1.2  
($\gamma_2$ is fully determined by $\gamma_1$). The calculated  
dependence of ${M'}^{0\nu}$ is shown in Fig.~\ref{0nbbfig:1}  
for $^{76}$Ge and $^{100}$Mo. One can see that over this range of  
$\gamma_1$ values the calculated results  
differ by only about 10\% from the ones corresponding to $\gamma_1=1.1$  
used in the standard calculations. 
 
Since  $M^{2\nu}$ is not affected by the short range repulsion,  
Fig.~\ref{0nbbfig:1} simply shows that the $0\nu\beta\beta$ 
matrix elements  ${M'}^{0\nu}$ are not very sensitive to reasonable 
variations of the parameter $\gamma_1$ \cite{foot2}. 
 
Combining the average $\langle {M'}^{0\nu} \rangle$ with the phase-space 
factors, the expected half-lives 
(for RQRPA and $\langle m_{\nu} \rangle$ = 50 meV, the scale 
of neutrino masses suggested by oscillation experiments) are also 
shown in Table \ref{tab:t12} (column 7).

\begin{figure}[t] 
  \begin{center} 
    \leavevmode 
    \epsfxsize=1.\textwidth 
    \epsffile{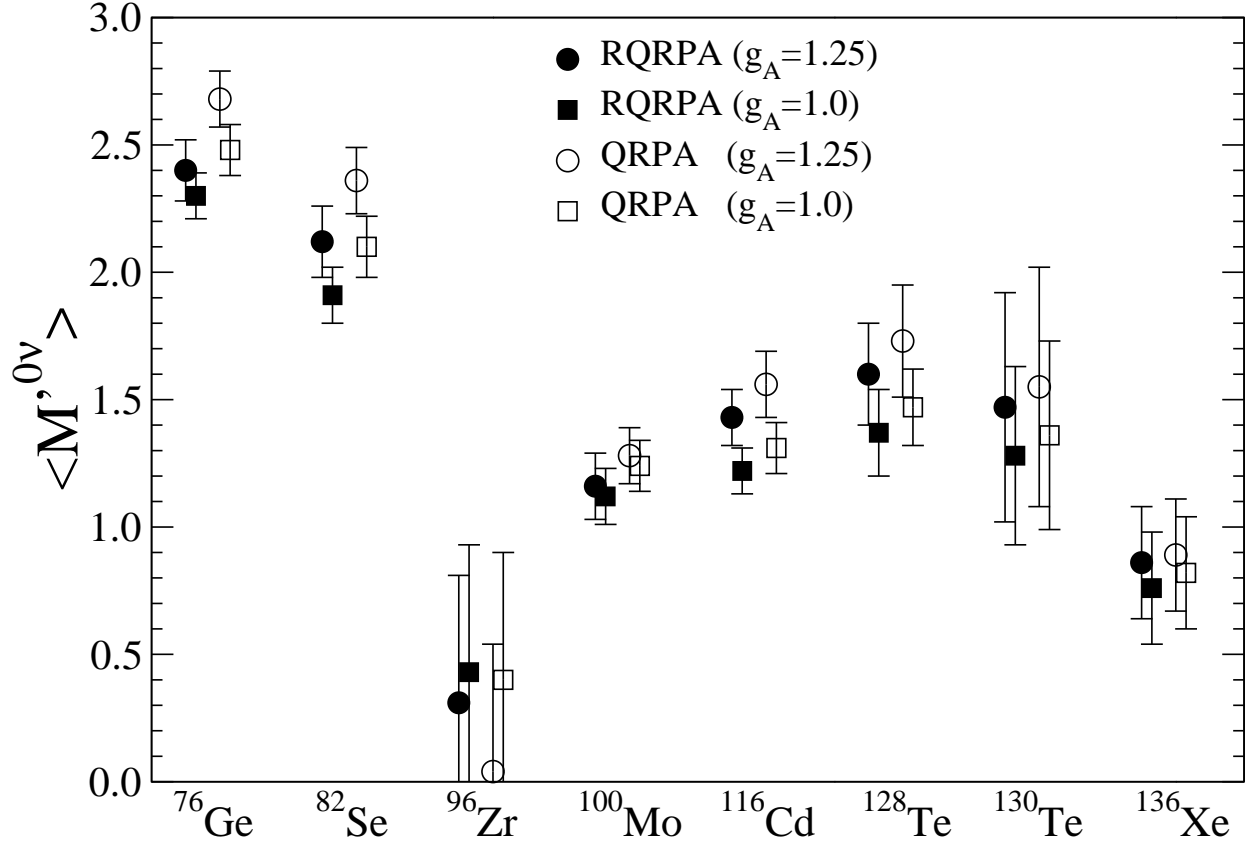} 
\vspace{0.5cm} 
    \caption{Average nuclear matrix elements $\langle {M'}^{0\nu} \rangle $ 
and their variance (including the error coming from the experimental uncertainty in $M^{2\nu}$) 
for both methods and for all considered nuclei. For $^{136}$Xe the error bars encompass the whole interval related 
to the unknown rate of the $2\nu\beta\beta$ decay.} 
    \label{0nbbfig:2} 
  \end{center} 
\end{figure}

The information collected in Table  \ref{tab:t12} is presented in graphical 
form in  Fig.~\ref{0nbbfig:2}. There 
the averaged nuclear matrix elements for both methods and both choices of $g_A$ are  
shown along with their full uncertainties (theoretical plus  
experimental). 
 
The entries for $^{130}$Te are slightly different from the corresponding results 
in Ref.\cite{Rod03a}. This is so because in the present work we use the $2\nu\beta\beta$ 
lifetime tentatively determined in the recent experiment \cite{Cuore}, while in 
\cite{Rod03a} we used the somewhat longer $2\nu\beta\beta$ lifetime based on the  
geochemical determination 
\cite{geoch}.

\subsection{Preliminary discussion of the procedure} 
 
\begin{figure}[t] 
  \begin{center} 
    \leavevmode 
    \epsfxsize=0.9\textwidth 
    \epsffile{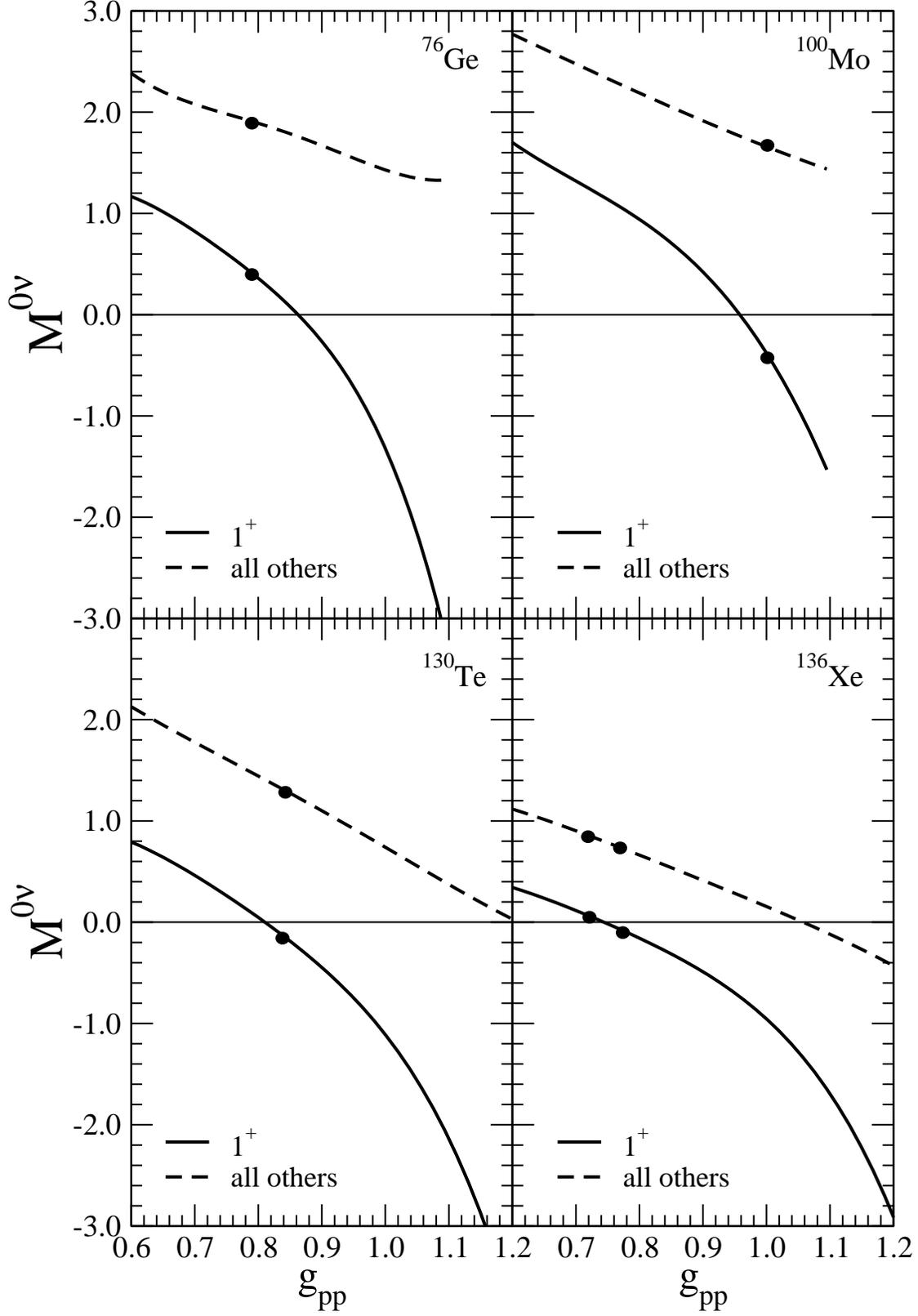} 
    \caption{Contribution of the $1^+$ and all the other (summed) multipoles to the  
matrix elements ${M'}^{0\nu}$ as a function of the parameter $g_{pp}$. 
The results were obtained within the RQRPA approach for the large model space. 
The dots indicate the $g_{pp}$ values that reproduce the known $2\nu$ decay rate. 
} 
    \label{0nbbfig:3} 
  \end{center} 
\end{figure}

\begin{figure}[t] 
  \begin{center} 
    \leavevmode 
    \epsfxsize=0.9\textwidth 
    \epsffile{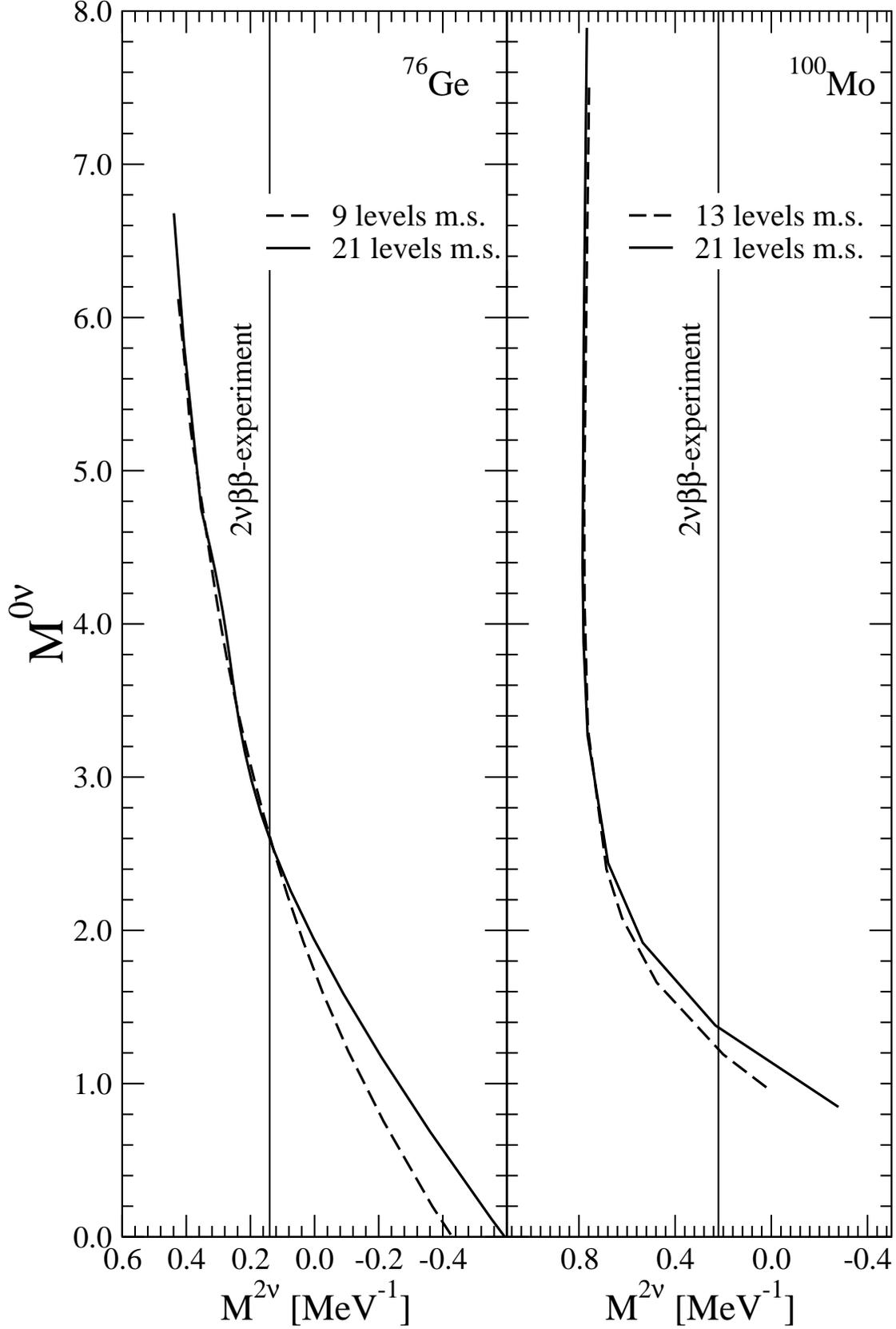} 
    \caption{Dependence of the matrix elements ${M'}^{0\nu}$ on  
$M^{2\nu}$ originating from the variation of the parameter $g_{pp}$ only. 
The results were obtained within the QRPA approach. The vertical lines show the
experimental $M^{2\nu}$ values used for the  $g_{pp}$ adjustment.
} 
    \label{0nbbfig:4} 
  \end{center} 
\end{figure}

\begin{figure}[t] 
  \begin{center} 
    \leavevmode 
    \epsfxsize=0.9\textwidth 
    \epsffile{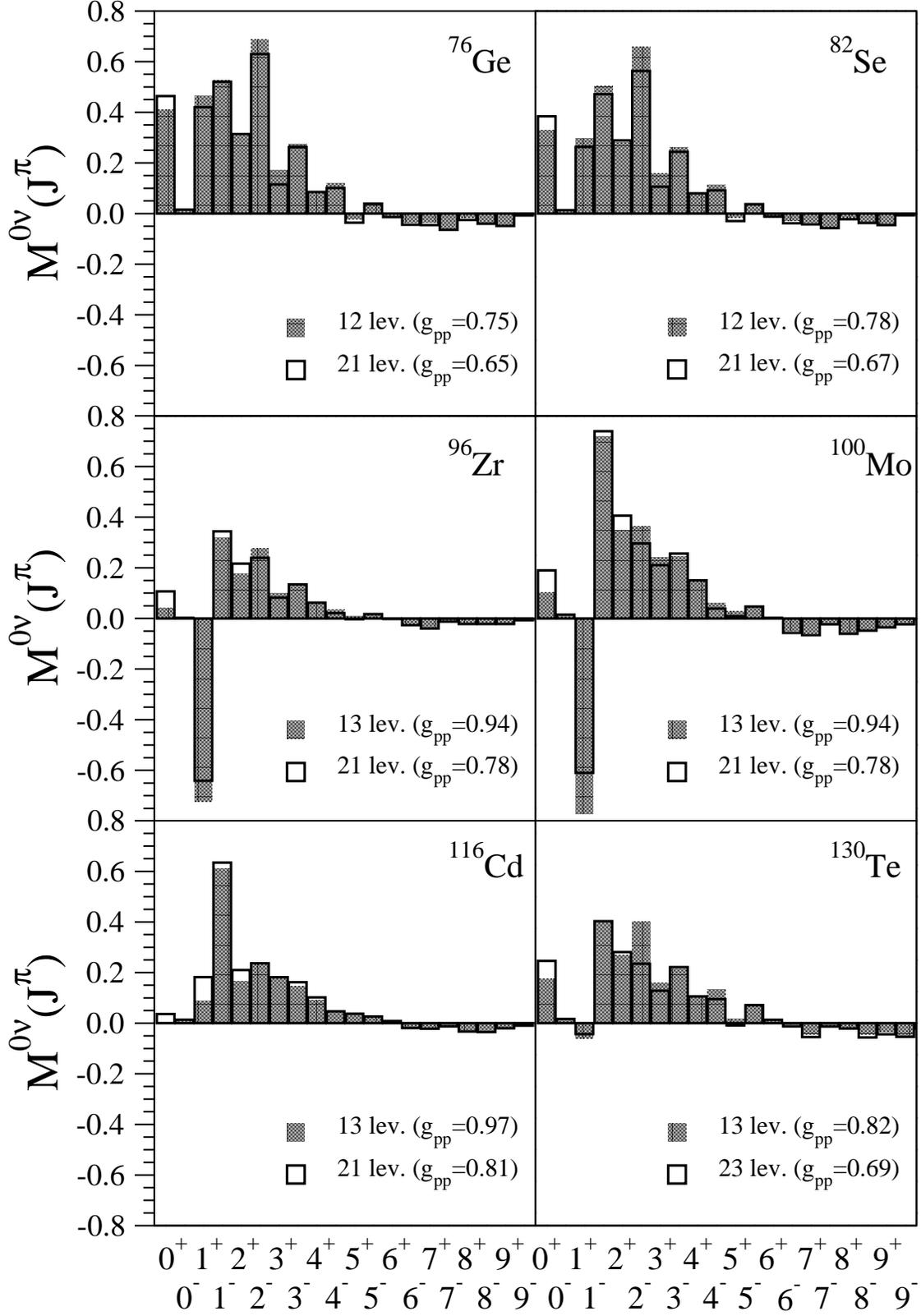} 
\vspace{0.5cm} 
    \caption{Multipole decomposition of the matrix elements ${M'}^{0\nu}$ 
calculated within the QRPA. $J^{\pi}$ is the angular momentum
and parity of the virtual intermediate state.}
    \label{0nbbfig:5} 
  \end{center} 
\end{figure}

One can qualitatively understand why our chosen procedure stabilizes 
the ${M'}^{0\nu}$ matrix elements as follows: The $M^{2\nu}$ matrix elements 
involve only the $1^+$ (virtual) states in the intermediate odd-odd 
nucleus. The nuclear interaction is such that the Gamow-Teller 
correlations (spin one, isospin zero pairs; after all the deuteron 
is bound and the di-neutron is not) are very near the 
corresponding phase-transition in the $1^+$ channel (corresponding 
to the collapse of the QRPA equations of motion). The contributions 
of the $1^+$ multipole for both modes of the $\beta\beta$ decay 
($2\nu$ and $0\nu$) depends therefore very sensitively on the 
strength of the particle-particle force, parameterized by $g_{pp}$. 
On the other hand,  
the  ${M'}^{0\nu}$ matrix element, due to the presence of the neutrino 
propagator, depends on states of many multipolarities in the virtual intermediate 
odd-odd nucleus. The other multipoles, other than $1^+$, correspond 
to small amplitudes of the collective motion; there is no instability 
for realistic values of $g_{pp} \sim 1.0$. Hence, 
they are much less sensitive to the value of $g_{pp}$. In the four panels 
of Fig. \ref{0nbbfig:3} we show the $g_{pp}$ dependence of the contribution of 
the $1^+$ multipole on one hand and of all the other multipoles added on the 
other hand. Remembering that the nominal value of $g_{pp}$ is $\sim 1.0$, 
one can clearly see the large difference in the corresponding slopes.    
By making sure that the contribution of the $1^+$ multipole is fixed, we  
therefore stabilize the 
${M'}^{0\nu}$  value. The fact that RQRPA essentially removes the 
instability becomes then almost irrelevant thanks to the  
chosen adjustment of $g_{pp}$.    
The effect of stabilization against the variation of the basis size is seen clearly  
if one plots directly ${M'}^{0\nu}$ versus $M^{2\nu}$, Fig.~\ref{0nbbfig:4}. 
Also the obtained multipole decompositions of ${M'}^{0\nu}$  
plotted in Fig.~\ref{0nbbfig:5} 
show the essential  
stability of the partial contributions against variation of the basis size. 
(Note that in the case of 
$^{96}Zr$ our approach gives rather small matrix element and its 
uncertainty is even  compatible with zero value. The smallness of that matrix element  
is due to a large negative contribution of the $1^+$ multipole,  
see Fig. \ref{0nbbfig:5}).

Since the $2\nu$ lifetime depends on the square of $M^{2\nu}$, there is an ambiguity
in choosing the sign of  the $M^{2\nu}$, and hence the corresponding value
of $g_{pp}$. In this work, and in Ref.\cite{Rod03a}, always the solution
corresponding to the smaller $g_{pp}$ value is chosen (positive $M^{2\nu}$
with the present phase convention). It is worthwhile to justify such a choice.
There are several reasons why the smaller $g_{pp}$ should be used.
First, QRPA and RQRPA are methods designed to describe small amplitude excitations
around the mean field minimum. Were we to choose the larger value
of $g_{pp}$, close or past the critical `collapse' value, the method
would be less likely to adequately describe the corresponding states.
Second, as we will show later, by choosing the larger $g_{pp}$ the disagreement
between the experimental and calculated rate of the single beta transitions
from the lowest $1^+$ state in the intermediate nucleus
would be far worse than with the smaller $g_{pp}$. Moreover, as shown
in Ref. \cite{deform}, only with the smaller $g_{pp}$ can one successfully
describe the systematic of single beta decay in a variety of nuclei.
It follows from the study of Ref. \cite{homma}
that choosing the larger value of $g_{pp}$ (i.e., the negative sign
of $M^{2\nu}$) would lead to a complete disagreement with the systematics
of single beta decays.
Finally, there is also a pragmatic argument for such a choice.
Only with it, the $0\nu$ nuclear matrix element becomes independent
of the size of the single particle basis. Thus, this choice,
admittedly ad hoc, removes the dependence on many more essentially
arbitrary choices. 
 
\begin{figure}[t] 
  \begin{center} 
    \leavevmode 
    \epsfxsize=0.95\textwidth 
    \epsffile{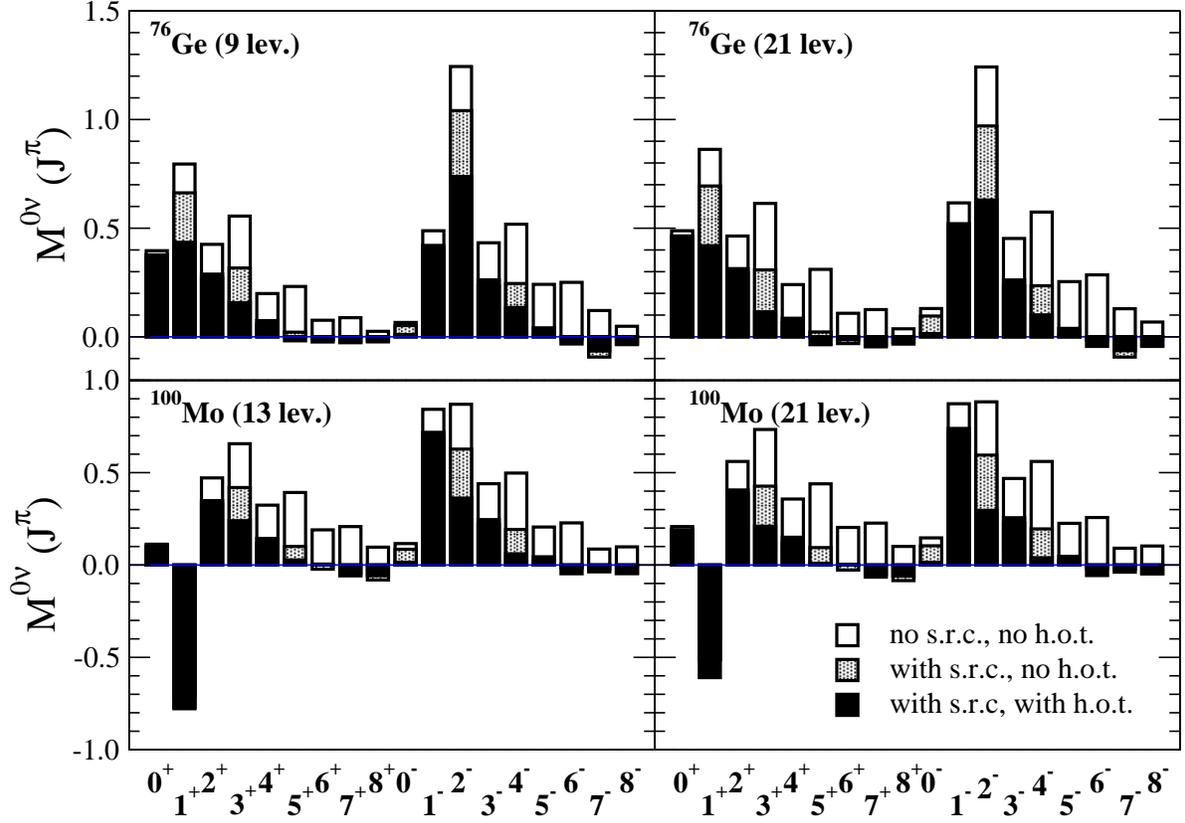} 
\vspace{0.3cm} 
    \caption{The effects of higher-order terms of nucleon currents (h.o.c.)
and of the nucleon-nucleon short range repulsion (s.r.c.) on the multipole 
distribution of the  $0\nu\beta\beta$-decay matrix element in the QRPA.
The left panels were calculated with 9(13) single-particle levels and
the right panels with 21 single-particle levels.
    \label{0nbbfig:6}} 
  \end{center} 
\end{figure}

One can see in Fig.\ref{0nbbfig:5} that all multipoles $J^{\pi}$  
(see Eq.(\ref{eq:long})), with the exception of the $1^+$ and various 
very small entries, contribute with the same sign. This suggests that 
uncertainties in one or few of them will have relatively minor effect. 
It is instructive to see separately the effect of the sometimes
neglected short-range repulsive nucleon-nucleon repulsion and of
the induced weak nucleon currents. These effects are shown in Fig.
\ref{0nbbfig:6}. One can see in that figure that the conclusion
of the relative role of different multipoles is affected by these
terms. For example, in $^{100}$Mo the $1^-$ multipole is the strongest 
one when all effects are included, while the $2^-$ becomes dominant
when they are neglected.

\begin{figure}[t] 
  \begin{center} 
    \leavevmode 
    \epsfxsize=0.82\textwidth 
    \epsffile{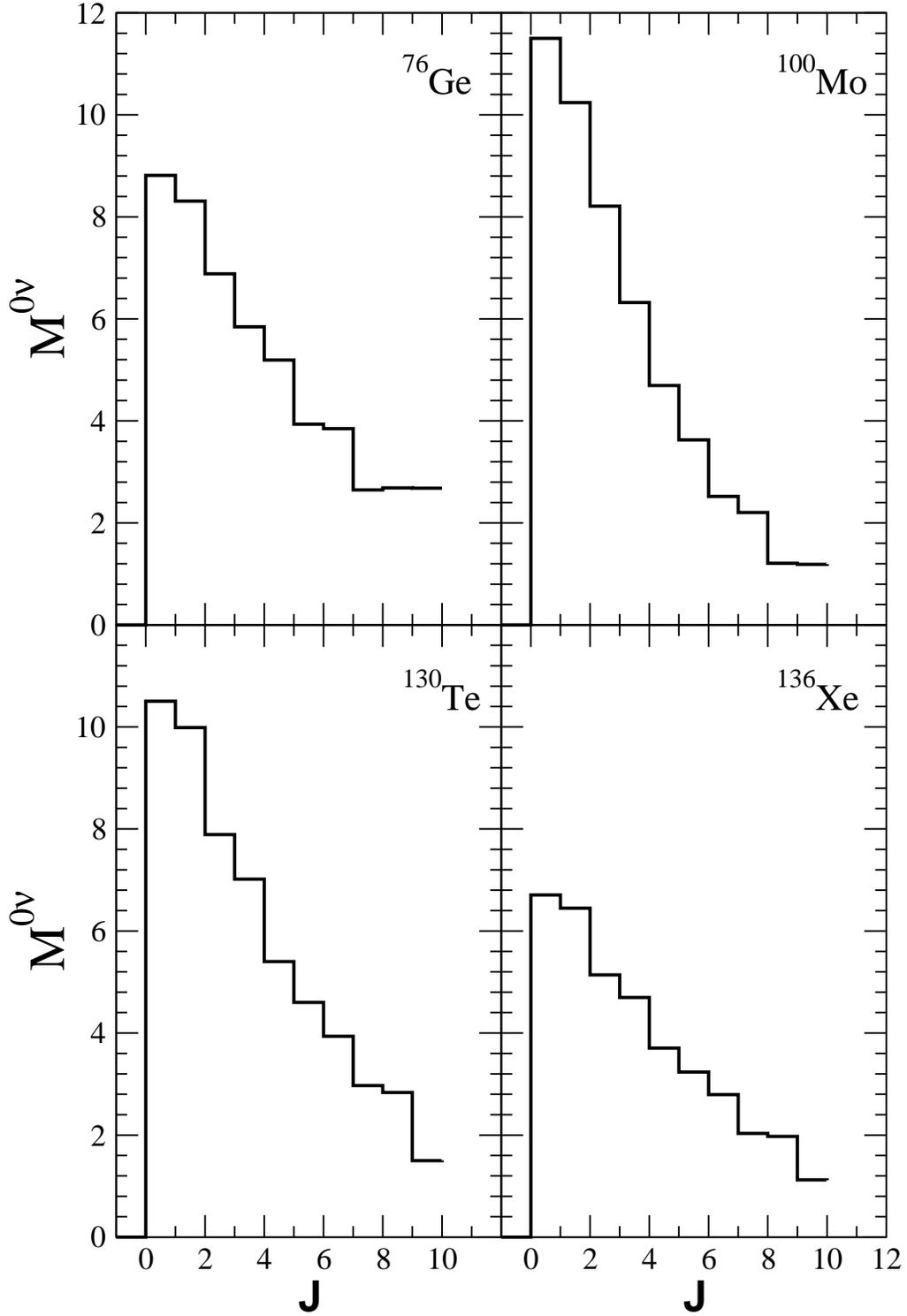} 
\vspace{0.3cm} 
    \caption{Running sum of the $0\nu\beta\beta$ decay 
matrix element as a function of the angular momentum ${\mathcal J}$ of the 
$nn$ and $pp$ pair that undergoes transition.} 
\label{0nbbfig:7} 
\end{center} 
\end{figure}

It is worthwhile to point out that one can display various contributions to 
${M'}^{0\nu}$ in an alternative, perhaps more revealing, way. Instead of 
$J^{\pi}$ that represents virtual states in the intermediate odd-odd nucleus, 
one can decompose the result in terms of ${\mathcal J}$, the angular 
momentum of the neutron-neutron (and simultaneously proton-proton) pair 
that undergoes the transition. In that case ${\mathcal J} = 0^+$ 
represents the `pairing only' contribution, while  ${\mathcal J} \ne 0^+$ 
come from the ground state correlations, i.e., contributions 
from higher seniority (broken pairs) states. In Fig. \ref{0nbbfig:7} 
we show that these higher seniority states contributions consistently have 
the tendency to cancel the `pairing only' piece. Thus, the final 
${M'}^{0\nu}$ is substantially less than the  ${\mathcal J} = 0^+$ 
part only, signifying the importance of describing the ground state 
contributions properly. (This tendency is a well known effect, 
for both modes of the $\beta\beta$ decay. It has been discussed, e.g. 
in Ref. \cite{gensen}.)

\section{Uncertainties of the $0\nu\beta\beta$ decay matrix elements} 
 
One cannot expect that the QRPA-like and shell model calculations will  
lead to identical results due to substantial differences between both  
approaches.  
(However, we point out below that the results of the present approach 
and the shell model results differ relatively little whenever a comparison 
is possible.) 
Our goal in this section is to  
show that a majority of differences among various 
QRPA-like calculations can be understood. In addition, we discuss the 
progress in the field and possible convergence of the QRPA results. 
Based on our analysis, we suggest that it is  not appropriate to treat all  
calculated $0\nu\beta\beta$-decay matrix elements at the same level,  
as it is commonly done (see e.g., \cite{bah04,CS03}), and to estimate their 
uncertainty based on their spread. 
 
Here we shall discuss the differences among different 
published QRPA and RQRPA results.  
We shall not consider the problem of the proton-neutron pairing in the 
double beta decay \cite{pan96} and the self-consistent RQRPA (SRQRPA) 
calculations \cite{bob01}. A systematic study of the $0\nu\beta\beta$-decay 
matrix elements within the SRQRPA will be discussed in a forthcoming publication, 
where we show that within the considered approach of fixing nuclear  
structure input \cite{Rod03a} the results agree well with those obtained  
within the QRPA and the RQRPA.

Representative examples of the nuclear matrix elements  
calculated by different authors within the 
QRPA and RQRPA for nuclei of experimental interest are collected 
in Table \ref{tab:comp}. In order to understand the 
differences between the entries, let us enumerate the main reasons  
leading to a spread of the published  
QRPA and RQRPA results: 
\begin{list}{}{\leftmargin=\parindent \rightmargin=0pt 
\itemindent=-\leftmargin \itemsep=-5pt} 
\item i) {\it The quasiparticle mean field}. Since the model space  
considered is finite, the pairing interactions have been adjusted 
to fit empirical pairing gaps based on the nuclear binding 
energies. This procedure is followed practically 
by all authors. However, it fails for closed proton  
($^{116}$Sn)  and neutron ($^{136}$Xe) shells. Some 
authors~\cite{CS03}  
modify the single particle energies (mostly 
spherical Woods-Saxon energies) in the vicinity of the Fermi 
surfaces to reproduce the low-energy quasiparticle spectra 
of the neighboring odd-mass nuclei.  
\item ii) {\it Many-body approximations}. The RQRPA goes beyond the QRPA 
by partially taking into account Pauli principle violation 
in evaluation of the bifermion commutators.  
Based on that, one might expect that the RQRPA is more accurate than the unrenormalized 
QRPA. Calculations within 
solvable models \cite{sim01} support this conjecture. 
\item iii) {\it Nucleon-nucleon interaction}. In the $0\nu\beta\beta$-decay 
calculations both schematic zero-range~\cite{Engel88} 
and realistic interactions were considered. In Ref.~\cite{mut89} 
$G$-matrix of the Paris potential approximated by a sum of Yukawa 
terms was used. The interaction employed by the Tuebingen group 
has been the Brueckner $G$ matrix that is a solution of the 
Bethe-Goldstone equation with Bonn (Bonn CD, Argonne, Nijmegen) one 
boson exchange potential. The results do not depend significantly 
on the choice of the NN interaction \cite{Rod03a}. 
\item iv) {\it The renormalization of the particle-hole interaction}. 
This is achieved by  scaling the particle-hole part of the (R)QRPA  
matrix by the parameter $g_{ph}$.   
That parameter  is typically adjusted by requiring that the energy  
of some chosen collective state, often the giant Gamow-Teller  
(GT) resonance, is correctly reproduced. In Ref. \cite{Rod03a} 
it was found that the GT state is almost independent of the size 
of the model space and is well reproduced with $g_{ph}\approx 1$. 
In Refs. \cite{pan96,si99} $g_{ph}=0.8$ was chosen and in Ref. 
\cite{suh91} $g_{ph}=1.3$ was considered for $A>114$ nuclei. 
The sensitivity of results to the change of $g_{ph}$ were 
studied in Ref.~\cite{bob01} where it was shown that for 
$0.8\le g_{ph}\le 1.0$ the change of 
the RQRPA $0\nu\beta\beta$-decay matrix element does not 
exceed 10\%. 
\item v) {\it The renormalization of the particle-particle interaction}. 
This is achieved by scaling the particle-particle part of the (R)QRPA  
matrix by the parameter $g_{pp}$. 
When the early QRPA calculation were performed,  
only a limited information about the experimental 
$2\nu\beta\beta$-decay half-lives 
was available. Thus, in Ref. \cite{Engel88} a probable $g_{pp}$ window 
was estimated from $\beta^+$ decays of semi-magic nuclei. In another 
work~\cite{mut89} systematics of the $pp$-force were investigated  
by an analysis of the single $\beta^+/EC$ decays. Alternatively, 
in many works $g_{pp}=1.0$ was chosen  \cite{si99,pan96,bob01,suh91,tom91}. 
Nowadays, the $2\nu\beta\beta$-decay has been observed in ten 
nuclides, including decays into two excited states \cite{data}.  
Two recent papers used the $2\nu\beta\beta$-decay half-lives to fix the strength 
of the particle-particle interaction of the nuclear Hamiltonian.  
In Ref. \cite{SK01} it was used only for the $J^\pi = 1^+$ channel leaving 
the particle-particle strength unrenormalized (i.e., $g_{pp} = 1.0$) 
in other channels. In our 
previous paper \cite{Rod03a}, and in the current work, $g_{pp}$ is adjusted so 
that the $2\nu\beta\beta$-decay rate is correctly reproduced. The same 
$g_{pp}$ is used for all multipole channels of the particle-particle 
interaction. Since information that can be used to adjust the $g_{pp}$
value exists only for the $1^+$ channel, we cannot make a separate
adjustment for other multipoles. Our economical choice is then to
use the same $g_{pp}$ for all multipoles. That makes the hamiltonian 
as simple as possible, and preserves the relative strength of different
multipoles of the realistic starting point interaction.
On the other hand,  
some authors prefer to fix the $g_{pp}$ value to the $\beta^-$ 
decay transition of the ground state of the intermediate nucleus 
\cite{CS03}. Such procedure can be carried out, however
only for three nuclear systems (A=100, 116 and 128) 
where $1^+$ is the ground state of the intermediate nucleus.
Since forbidden decays are less well understood than the allowed 
ones it is difficult to rely on them
in the case of other nuclear systems. 
(We discuss in more detail the differences between these two approaches of  
fixing the parameter $g_{pp}$ in the next section.)  
\item vi) {\it The size of the model space.} A small model 
space comprising usually of two major shells was often used in the 
calculation of the nuclear matrix elements  
\cite{CS03,SK01,Engel88,mut89,suh91,tom91}. Significantly 
larger model space has been used in Refs.\cite{Rod03a,si99} consisting 
of five major shells. As a rule the results obtained for the large model space 
are reduced when compared to those for the small  
model space for the same value of  
$g_{pp}\sim 1$. The dependence of the calculated nuclear matrix elements on the 
size of the model space has been studied in Ref.\cite{SK01}  
and in \cite{Rod03a}. Rather different conclusions using the same procedure 
of fixing $g_{pp}$ was reached. The origin of such  
discrepancy is not understood. In Ref. \cite{SK01} relatively small differences 
between model spaces were considered and a large effect on the  
$0\nu\beta\beta$-decay  matrix element was found. In particular, 
by adding the $N = 2$ shell to the $N = 3-4$  
oscillator shell model space the $0\nu\beta\beta$-decay matrix element 
of $^{76}Ge$ decreased by a factor of about 3 (see Table \ref{tab:comp}). 
Thus, in Ref. \cite{SK01} the single-particle levels lying 
far from the Fermi surface seem to influence strongly the decay rate. 
In contrast, in Ref. \cite{Rod03a} and in the present work results
essentially independent on the size of the basis  
have been found 
even with significantly different sizes of the model space. 
\item vii) {\it The closure approximation.} 
The $0\nu\beta\beta$-decay matrix elements were usually calculated 
using the closure approximation for intermediate nuclear states  
\cite{SK01,Engel88,mut89,tom91}.  
Within this approximation energies of intermediate states ($E_n-E_i$) are replaced 
by an average value ($<E_n-E_i>\approx 10~MeV$), and the sum over intermediate 
states is taken by closure, $\sum_n |J^\pi_n><J^\pi_n| = 1$. This simplifies 
the numerical calculation drastically. The calculations with exact treatment 
of the energies of the intermediate nucleus were presented in Ref.  
\cite{Rod03a,si99,pan96}. The effect of the closure approximation was 
studied in details in Ref. \cite{mut94}. It was found that the  
differences in nuclear matrix elements are within $10\%$. This is so because  
the virtual neutrino has an average momentum of  $\sim 100$ MeV, 
considerably larger than the differences in nuclear excitation 
energies. 
\item viii) {\it The axial-vector coupling constant $g_A$.} 
The axial-vector coupling constant
or in other words, the treatment of quenching, is also a source of differences 
in the calculated nuclear matrix elements. 
The commonly adopted values are 
$g_A=1.0$ \cite{Engel88} and $g_A=1.25$ \cite{Rod03a,si99,pan96,mut89,tom91}. 
However, as shown in  Table \ref{tab:comp}, if $g_{pp}$ is fixed to the 
$2\nu\beta\beta$-decay half-life, the effect of $g_A$ 
modification is smaller, of order $10\%$. 
\item ix) {\it The two-nucleon short-range correlations (s.r.c.)}. 
In majority of calculations the short-range correlations between  
two nucleons are taken into account by multiplying the two-particle wave functions  
by the correlation function \cite{MS1976}: 
$f(r)=1-e^{-\gamma_1 r^2}(1-\gamma_2 r^2)$  
with $\gamma_1 = 1.1$~fm$^{-2}$ and $\gamma_2 = 0.68$~fm$^{-2}$. 
The sensitivity of results to the change of this parameters have been discussed 
in the previous section. It is also known that the $0\nu\beta\beta$-decay 
matrix element is more affected (reduced) by the s.r.c. at higher $g_{pp}$
values. 
We note that the s.r.c. were not taken into account within the approach which
was developed in Ref.\cite{suh91} and used in the recent
publication~\cite{CS03}. For the realistic $g_{pp}$ values neglecting
s.r.c. would lead to an increase in $M^{0\nu}$ by factor of about 2.
\item x) {\it The higher order terms of the nucleon current.} 
The momentum dependent higher order terms of nucleon current, namely the 
induced pseudoscalar and weak magnetism, were first considered 
in connection with the light neutrino mass mechanism of the $0\nu\beta\beta$-decay in 
\cite{si99}. Their importance is due to the virtual character of the exchanged 
neutrino, with a large average momentum of $\sim$ 100 MeV.  
The corresponding transition operators have  
different radial dependence than the traditional ones. It is worth mentioning  
that with modification of the nucleon current one gets a new 
contribution to neutrino mass mechanism, namely the tensor contribution. 
The corrections due to the induced pseudoscalar nucleon 
current vary slightly from nucleus to nucleus, and in all cases represent about 
$30\%$ reduction.  In the present calculation the effect of higher  
order terms of nucleon current is taken into account. 
\item xi) {\it The finite size of the nucleon} is taken into account via  
momentum dependence of the nucleon form-factors. Usually this effect  
on the $0\nu\beta\beta$-decay matrix elements  
associated with light neutrino exchange is neglected \cite{Engel88,tom91}, 
since it is expected to be small. In the presented calculation it is taken account 
by assuming phenomenological cutoff (see \cite{si99} and references therein). 
We found that by considering the finite nucleon size the value of ${M'}^{0\nu}$ 
is reduced by about 10\%. In Refs. \cite{suh91,CS03} the nucleon current 
is treated in a different way than in other double beta decay studies, namely 
it is evaluated from the quark level using relativistic quark wave  
functions. However, the momentum dependence of corresponding  
nucleon form-factors is not shown, so a comparison is difficult.    
Comparing the present results with those of Ref. \cite{suh91} for 
the same nuclear structure input suggests that the agreement might be  
achieved only if a very low cutoff is introduced. 
\item xii) {\it The overlap factor of intermediate nuclear states}. 
This factor is introduced since the two sets of intermediate 
nuclear states generated from initial and final ground states are 
not identical within the considered approximation scheme. A majority  
of calculations uses a simple overlap factor $X^iX^f-Y^iY^f$,  
inspired by the orthogonality condition of 
RPA states generated from the same nucleus. However, the double beta 
decay is a two-vacua problem. The derivation of the overlap 
matrix within the quasiboson approximation scheme was performed 
in Refs.\cite{overl,deform}. It was shown that the overlap factor 
of the initial and final BCS vacua is an integral part of the overlap 
factor of the intermediate nuclear states \cite{deform,grotz}. This BCS overlap 
factor, about 0.8 for spherical nuclei,  
is commonly neglected in the calculation of nuclear matrix elements.  
If one fits $g_{pp}$ to the experimental $2\nu\beta\beta$-transition probability,
the effect of the BCS overlap factor is significantly reduced. Hence it
is neglected in the present work.  
This factor has been found to be important in the calculation 
within the deformed QRPA. 
\item xiii) {\it The nuclear shape.} Until now, in all QRPA-like calculations of the 
$0\nu\beta\beta$-decay matrix elements the spherical symmetry was assumed 
as the majority  of nuclei of experimental interest are nearly  
spherical. The effect of deformation on the $0\nu\beta\beta$-decay matrix 
elements has not been studied as of now. Recently, the $2\nu\beta\beta$-decay matrix  
elements were calculated within the  
deformed QRPA with schematic forces \cite{deform}.  
It was found that differences in deformation between initial and final nuclei   
have a large effect on the $2\nu\beta\beta$-decay half-life. One could expect 
that a similar mechanism of suppression of nuclear matrix elements is 
present also in the case of the $0\nu\beta\beta$-decay. It goes without  
saying that a further progress concerning this topics  is highly desirable.  
The deformed QRPA might be the method of choice  for the description of  
double beta decay of heavy nuclei like $^{150}$Nd, or $^{160}$Gd. However,  
the deformation of many medium heavy nuclear systems are also noticeable  
(see Table II of \cite{deform}). The main difficulty of the deformed RQRPA  
calculation is the fact that  by going from the spherical to the deformed 
nuclear shapes  
the number of configurations increases drastically. 
\end{list} 
 
The present Table \ref{tab:comp},  
is organized as follows: 
i) The results with and without inclusion of higher order terms 
of nucleon current are separated. We point out that in Refs.  
\cite{CS03,suh91,au98} the contribution from the weak-magnetism 
was taken into account. It is, however, negligible and we present 
their results without this contribution. ii) We indicate those 
matrix elements, which differ significantly from the present ones or 
other previous 
calculations (denoted as SK-01 \cite{SK01} and CS-03 \cite{CS03}). 
We shall discuss this problem in detail below. iii)  
Within the sub-blocks the nuclear matrix elements  are presented  
in ascending order following the year they appeared. 
 
Many  nuclear matrix elements in Table \ref{tab:comp} 
were calculated from the published ratios of Fermi $M^{0\nu}_F$  
and Gamow-Teller $M^{0\nu}_{GT}$ contributions and  
the absolute values of $|M^{0\nu}_{GT}|$ presented in units of fm$^{-1}$, 
i.e., not scaled with nuclear radius \cite{pan96,suh91,tom91,au98} 
We assumed that $R = r_0 A^{1/3}$ with $r_0 = 1.1$ fm.  
However, Ref. \cite{mut89} used $r_0=1.2$ fm, so we rescaled the 
results in order to compare them better with ours.  
The differences between many calculations are understandable just 
from the way $g_{pp}$ was fixed, the considered size of the model  
space, the inclusion of the short-range correlation, the way 
the finite nucleon size is taken into account and other minor effects. 
Some nuclei, like $^{100}$Mo, exhibit more sensitivity to these  
effects than others, such as $^{76}$Ge or $^{82}$Se. That is confirmed
by our numerical studies.  
 
The calculations of the $0\nu\beta\beta$-decay matrix elements  
by Civitarese and Suhonen \cite{CS03} (denoted as CS-03 in Table  
\ref{tab:comp}) deserves more comments. The authors performed them  
within the approach suggested  
in Ref. \cite{suh91}, (SKF-91) which employs the nucleon  
current derived from the quark wave functions. In this approach the two  
nucleon short-range-correlations are not taken into account. 
But, the contribution from  higher order terms of nucleon current 
is studied. Unlike the present results
the authors of Ref.\cite{suh91} claim 
that contribution from the induced pseudoscalar coupling  
is a minor one and they do not include it. 
At the same time, they find that 
the contribution from the weak-magnetism is negligible \cite{suh91}, 
in agreement with our result and with the $1/m_p^2$ scaling. 
 
In Ref. \cite{suh91} the $g_{pp}$  
was taken to be unity. The extension of this work for the case  
when $g_{pp}$ is adjusted to reproduce the 
single $\beta$-decay decay amplitudes was presented 
in Ref.\cite{au98} (AS-98) where also the  
effect of adjustment of the single particle energies to agree 
better with the spectroscopic data of odd-mass nuclei ($AS_{AWS}-98$)
was studied.  
Since the adopted value of $g_A$ used
for fixing $g_{pp}$ was not given, in Table \ref{tab:comp} we present
the corresponding ${M'}^{0\nu}$ for both $g_A = 1.0$ and $g_A = 1.25$.
(From the related article \cite{Suh04} it seems $g_A = 1.0$ was considered.)

In Ref. \cite{CS03} (CS-03) the  
nuclear matrix elements are calculated in the same way as in 
\cite{au98}, however, the obtained results differ significantly 
(see Table \ref{tab:comp}) from each other. For some 
nuclei the difference  is as large as a factor of two. There is 
no discussion of this there or in the later Refs. \cite{CS03,Suh04}. 
It is worth noticing that the largest matrix element in \cite{CS03} is 
found for the $0\nu\beta\beta$-decay of $^{136}$Xe. This 
disagrees with the results of other authors 
(see Table \ref{tab:comp}). The reduction of the $0\nu\beta\beta$-decay 
of the $^{136}$Xe is explained by the closed neutron shell  
for this nucleus. A sharper Fermi surface leads to a reduction 
of this transition. 
 
Altogether, the matrix elements of Refs.\cite{au98,CS03} are noticeably larger
than the present ones. Most of that difference can be attributed to the neglect of the 
short range nucleon-nucleon repulsion and of the higher order effects in the nucleon
weak current in these papers.

As the above list of 13 points shows, there are many reasons,  
some more important than others, 
that might cause a difference between various calculated $0\nu\beta\beta$ 
matrix elements. Clearly, when some authors do not include effects that 
should be included (e.g. the short range correlations or the higher order 
terms in the nucleon current) their results should be either corrected or 
convincing arguments should be given why the chosen procedure
was adopted. 
Other effects on the list are correlated, like the size of the model space 
and the renormalization of the particle-particle interaction. Again, 
if those correlations are not taken into account, erroneous conclusion might 
be drawn. Yet other effects are open to debate, like the quenching of $g_A$ 
or the adopted method of adjusting $g_{pp}$. In our previous work \cite{Rod03a}, 
and in the present one, we show that our chosen way of renormalization 
removes, or at least greatly reduces, the dependence of the final result 
on most of the effects enumerated above. 
 
Our choice of $g_{pp}$ adjustment, to the experimental $2\nu\beta\beta$ rate 
was used also in Refs.\cite{SK01,Muto97}. 
Let us comment on their results. Note, that the higher order corrections  
(h.o.c.) to the weak nucleon current were neglected in those works, hence one 
expects a $\sim$ 30\% discrepancy right away. 
 
In Ref. \cite{Muto97} only the  $\beta\beta$ decay of $^{76}$Ge was considered  
within both the QRPA and RQRPA. Only one s.p. basis was used,  
corresponding to the small one in our notation.  
The result was $M^{0\nu}=4.59\ (3.88)$ within QRPA (RQRPA).   
If one takes into account that h.o.c. as a rule reduce $M^{0\nu}$ by about 30\% and the 
fact that $\sim$ 10\% larger nuclear radius was used, one concludes that 
these matrix elements should be multiplied by $\sim$ 0.6 in order to compare 
them  with our calculations. This results in 
$M^{0\nu}=2.76\ (2.34)$ which are in a very good agreement  
with our calculated $M^{0\nu}=2.68\ (2.41)$ for this case. 
Hence QRPA and RQRPA gave quite similar results in that case. 
 
In contrast, the conclusions in Ref. \cite{SK01} are quite different from ours; 
the authors found significant dependence of their results on both the nuclear 
model and on the s.p. basis size, in spite of the adjustment of $g_{pp}$.  
The reason for these differences is unknown. In an attempt to understand
the origin of them we comment here on features that, in our judgment,
require further discussion. 
 
1) We have already pointed out that the $M^{0\nu}$ for $^{76}$Ge 
in Ref.  \cite{SK01} changes with 
changing the single-particle basis within QRPA -  
from 4.45 (9 levels) to 1.71 (12 levels),  
or more than 2.5 times. This was obtained by adding just the deep-lying $sd$ shell,
a surprising result.  
 
2) The last of the three papers contains most information 
and we can use their  Figs. 1-14 with $M^{2\nu}$ and Figs. 15-25 with $M^{0\nu}$,  
in order to compare the numbers in the first (QRPA) and the last (SQRPA) columns of their 
Table 3. We were able to reproduce the entries in the QRPA column, 
but almost all entries in the SQRPA column, apart from those for $^{82}$Se and $^{76}Ge$, 
seem inconsistent.  
The most striking difference is for $^{136}$Xe. By inspecting Fig. 7 in  \cite{SK01}, one  
finds that the appropriate $g_{pp}$ for SQRPA is about 1.08 (lines a and b). 
Now, by going to Fig. 24 one can see that the corresponding values of 
$M^{0\nu}$ are about 2.5 (lines c and d) while the entries in  Table 3 
are 0.98 and 1.03. 
 
3) Some of the results in  \cite{SK01} (see  Fig.1, lines a and b, SQRPA; Fig.2, 
lines a and b, SQRPA, Fig.5, lines c and d, QRPA), Fig.7, lines c and d, QRPA)  
contradict the generally accepted conclusion that  
for all RPA-like approaches the functions $M^{2\nu}(g_{pp})$ calculated with the larger  
basis cross zero faster than the ones obtained with the smaller one. 
Thus, given these apparent inconsistencies,  
it is difficult to draw any conclusions from the comparison of our 
and Ref.~\cite{SK01} results. 
    
As pointed out above, one cannot expect a perfect agreement between the 
present result and the large scale shell model results \cite{SM96}. The approximations 
are different, and the shell model calculations do not include 
several multipoles (which typically 
enhance the matrix elements, see Fig. \ref{0nbbfig:5}). Yet, remarkably, the 
present results and the published large scale shell model results agree 
with each other 
considerably better than the various entries in Table \ref{tab:comp}. 
So, using the published values, \cite{SM99}, we find ${M'}^{0\nu}$ values 
1.5, 2.1, 1.1, and 0.7 for $^{76}$Ge, $^{82}$Se, $^{130}$Te and $^{136}$Xe, 
while RQRPA gives 2.4, 2.1, 1.5, and 0.7-1.0 for the same nuclei. However, 
it appears that the more recent shell model results \cite{SM04} for  
$^{130}$Te and $^{136}$Xe give larger values while at the same time 
overestimating to some degree the rate of the $2\nu\beta\beta$ decay. 
In any case, we find this comparison encouraging. 
 
\section{Further discussion of the parameter adjustment} 
 
Ideally, the chosen nuclear structure method should describe 
all, or at least very many, experimental data and do that without 
adjustments. As described above that is not the case of QRPA 
or RQRPA. The interaction used is an effective interaction, 
and various parameters ($g_{pair}, g_{ph}, g_{pp}$) are adjusted.
In particular, the parameters $g_{pair}$ and $g_{pp}$
are adjusted on the case-by-case basis in the present approach.
Even then the method is not able to describe well all relevant 
weak transitions. In particular, it is sometimes impossible to describe 
simultaneously the $2\nu\beta\beta$ decay rate as well as 
the $\beta^-$ and $\beta^+/EC$ matrix elements connecting the 
$1^+$ ground states of the intermediate nucleus  
with the ground states of the final and initial nuclei 
($^{100}$Mo is a well known example of this problem, see e.g.~\cite{grif92}). 
 
Empirically, the transitions through 
the $1^+$ ground state of $^{100}$Tc, $^{116}$In, and $^{128}$I 
seemingly account for most of the $2\nu\beta\beta$ matrix element 
(this is the so called `Single State Dominance' \cite{ssd}).  
Thus it appears that 
these single $\beta$ transitions are particularly relevant. 
Based on such considerations, Suhonen \cite{Suh04} suggested 
that the $\beta^-$ matrix element is more suitable source 
for the $g_{pp}$ adjustment  than the $2\nu\beta\beta$ decay. 
Below we explain why we prefer the chosen method of
parameter adjustment. On the other hand, we also
argue that the adjustment based on the $\beta^-$ transitions
might give results not far from ours, provided the independence
on other adjustments can be proved.   
 
\begin{figure}[t] 
  \begin{center} 
    \leavevmode 
    \epsfxsize=0.82\textwidth 
    \epsffile{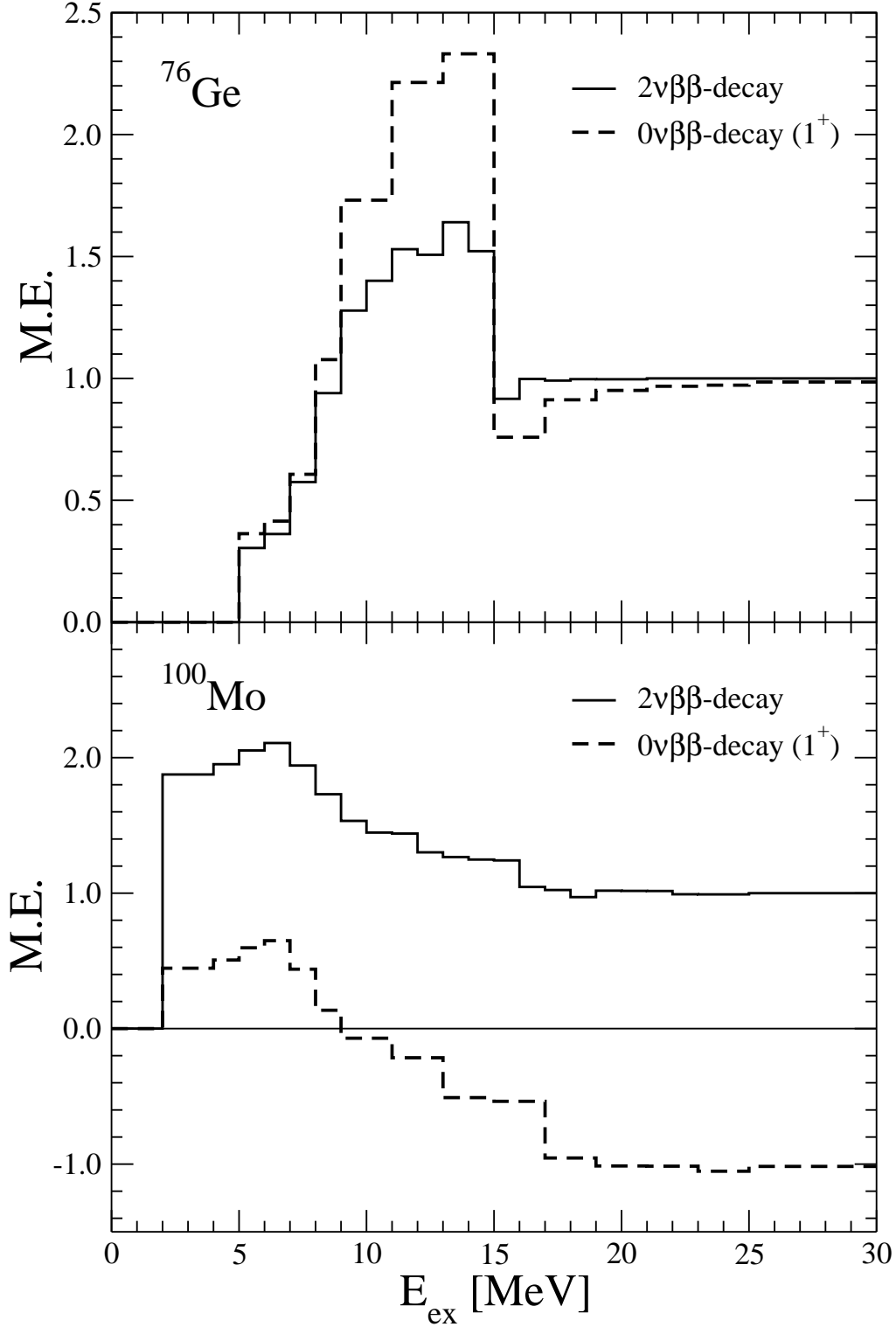} 
\vspace{0.3cm} 
    \caption{ 
Running sum of the $2\nu\beta\beta$-decay  
and $0\nu\beta\beta$-decay (only $1^+$ component) 
matrix elements for $^{76}Ge$ and $^{100}Mo$ (normalized to 
unity) as a function of the  
excitation energy $E_{ex}=E_n-(E_i+E_f)/2$. 
Calculations were performed within the QRPA. 
} 
    \label{0nbbfig:8} 
  \end{center} 
\end{figure}

First, it is not really true that the first $1^+$ state 
is the only one responsible for the $2\nu\beta\beta$ decay. 
This is illustrated for the cases of $^{76}$Ge and $^{100}$Mo 
in Fig. \ref{0nbbfig:8}. Even though for $^{100}$Mo 
the first state contributes substantially, higher lying states 
give non-negligible contribution. And in $^{76}$Ge many 
$1^+$ states give comparable contribution. Thus,  
to give preference to the lowest state is not
well justified, the sum is  
actually what matters. At the same time, the dilemma that 
the $\beta^-$ and   $\beta^+/EC$ matrix elements move with  
$g_{pp}$ in opposite directions makes it difficult to  
choose one of them.  
It seems better to use the sum of the products of the 
amplitudes, i.e. the  $2\nu\beta\beta$ decay. 
 
At the same time, the contribution of the $1^+$ multipole to the 
 $0\nu\beta\beta$ matrix element and the corresponding  
 $2\nu\beta\beta$ matrix element are correlated, even though they are 
not identical, as shown in Fig. \ref{0nbbfig:8}. Making sure that 
the  $2\nu\beta\beta$ matrix element agrees with its experimental value 
constrains the $1^+$ part of the  $0\nu\beta\beta$ matrix element  
as well. 
 
\begin{figure}[t] 
  \begin{center} 
    \leavevmode 
    \epsfxsize=0.85\textwidth 
    \epsffile{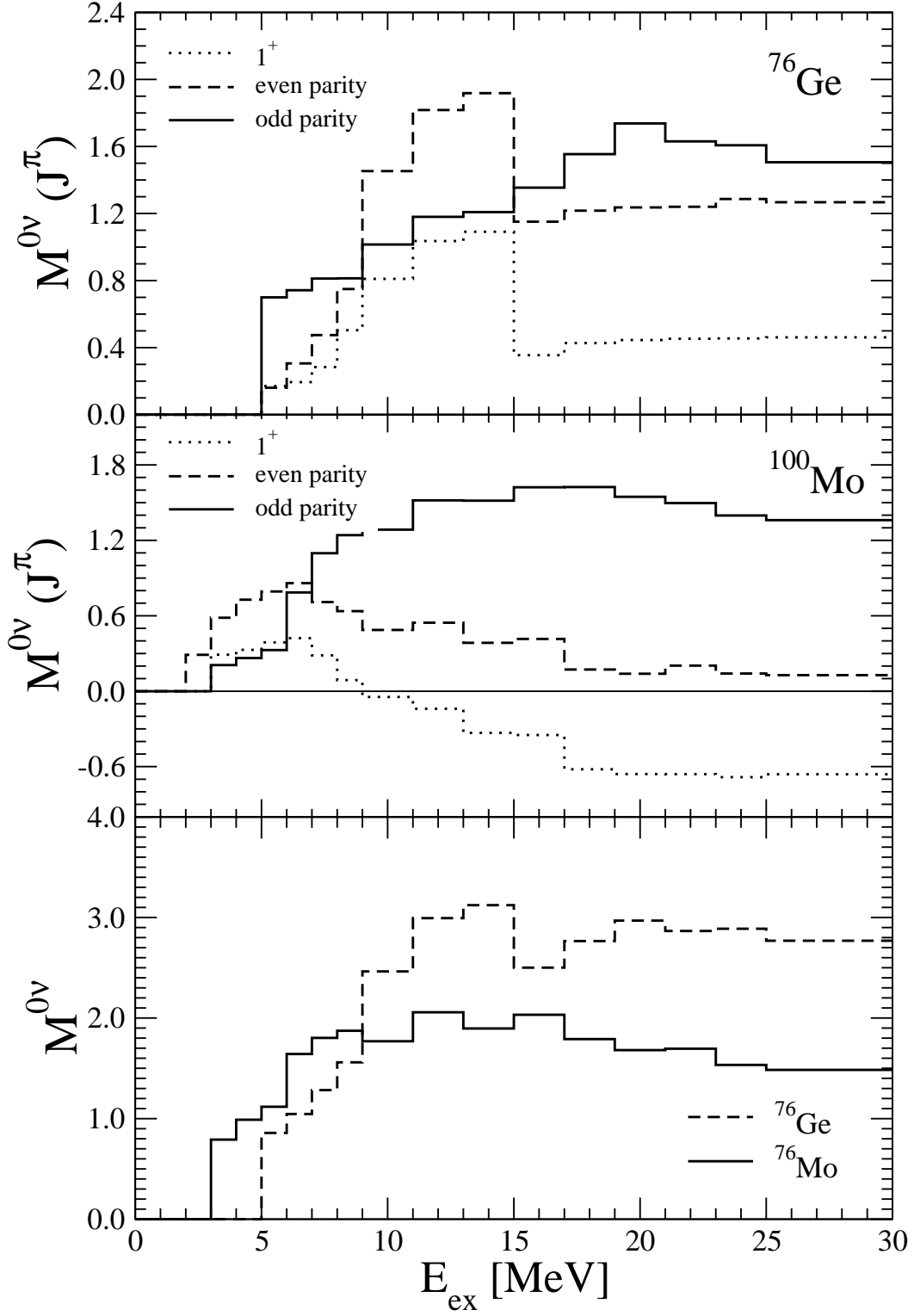} 
\vspace{0.5cm} 
    \caption{ 
Running sum of the $0\nu\beta\beta$-decay matrix element for $^{76}Ge$ 
and $^{100}Mo$ and their multipole contributions 
as a function of excitation energy $E_{ex}=E_n-(E_i+E_f)/2$. 
Calculations were performed within the QRPA (small model space). 
} 
    \label{0nbbfig:9} 
  \end{center} 
\end{figure}

In Fig. \ref{0nbbfig:9} we show the running sum contributions 
to the $0\nu\beta\beta$ matrix elements in  $^{76}$Ge and $^{100}$Mo, 
separated into multipoles, and for the total. Such a sum, even 
for the $1^+$ component, is rather different that the similar 
staircase for the $2\nu\beta\beta$; there is no single state 
dominance. Thus, again, it is not obvious that it is best to choose any one 
particular state or transition for the adjustment. 
Moreover, while we have demonstrated that adjusting $g_{pp}$ 
to the $2\nu\beta\beta$ rate removes the dependence on other 
parameters, a similar proof was not given in \cite{Suh04}. 
Finally, as seen in Fig. \ref{0nbbfig:3} the slope  
of the $1^+$ multipole component of the $0\nu\beta\beta$  
matrix element far exceeds the slope of the other multipoles.  
Hence, making sure that the whole $1^+$ multipole contribution 
is correct is crucial. 
 
Thus, we  
prefer the method of adjustment used in the present work.  
That is so not only for the reasons shown above, but also 
because it is much more general as stressed already, while the ground state 
of the odd-odd nucleus has $1^+$ only in  
$^{100}$Tc, $^{116}$In, and $^{128}$I. 

\begin{figure}[t] 
  \begin{center} 
    \leavevmode 
    \epsfxsize=0.79\textwidth 
    \epsffile{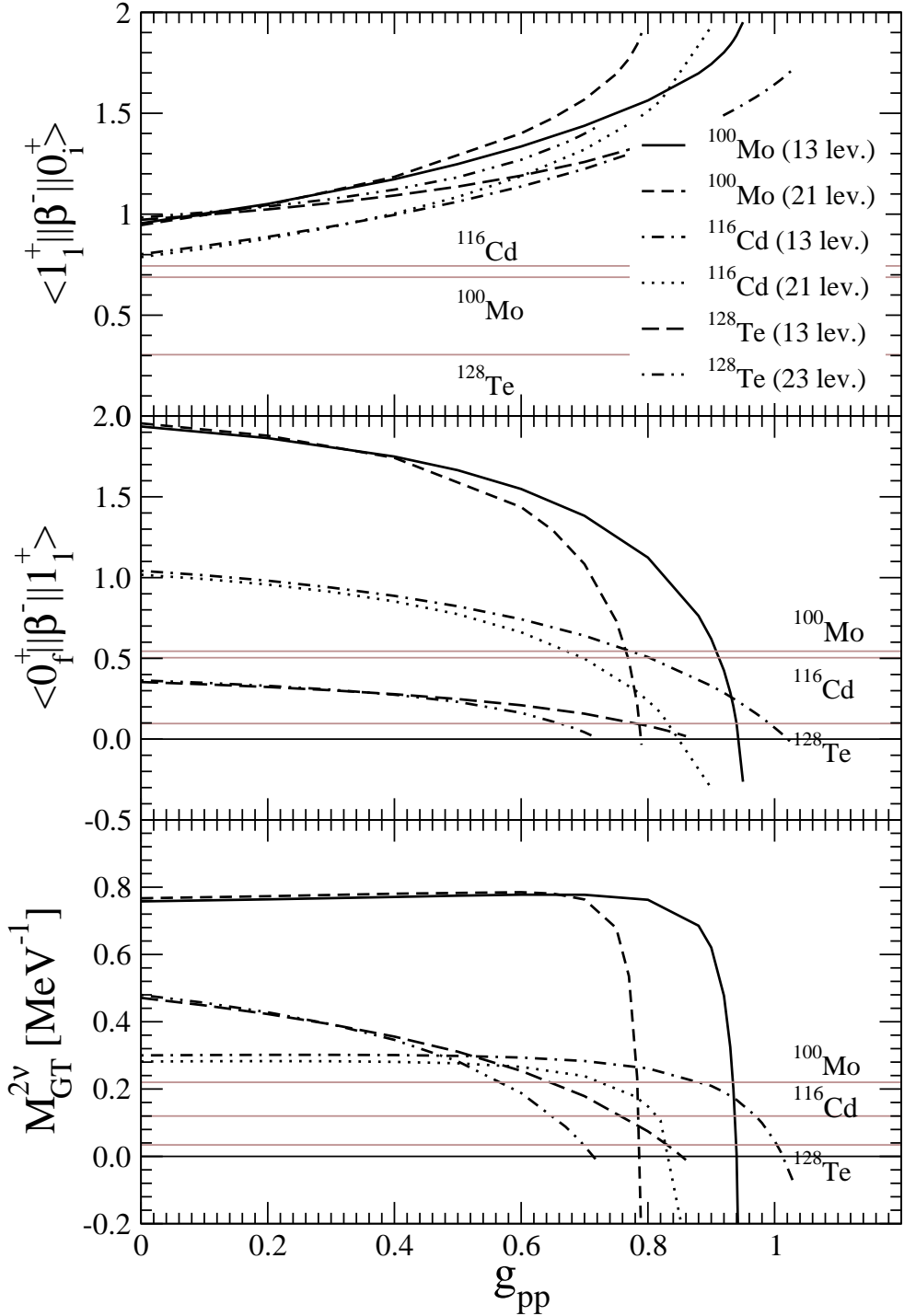} 
\vspace{0.3cm} 
    \caption{The upper panel: The matrix element  
associated with the $\beta^-$ transition from the initial nucleus 
(A,Z) to the $1^+$ ground state of the intermediate  
nucleus (A,Z+1) as function of $g_{pp}$ for $^{100}Mo$, 
$^{116}Cd$ and $^{128}Te$. 
The middle panel: The same as upper panel for the  
$\beta^-$ transition from the ground state of (A,Z+1) 
to ground state of final nucleus (A,Z+2). 
The lower panel: The $2\nu\beta\beta$-decay matrix element of  
$^{100}Mo$, $^{116}Cd$ and $^{128}Te$ as a function of $g_{pp}$. 
The experimental values are indicated by thin horizontal lines
($g_A=1.25$ is considered).} 
    \label{0nbbfig:10} 
  \end{center} 
\end{figure}

However, as demonstrated 
in the last figure, Fig.~\ref{0nbbfig:10}, the adjustment proposed in 
Ref. \cite{Suh04}, namely to choose the parameter $g_{pp}$ based 
on the $\beta^-$ decay of the intermediate $1^+$ state, would  
not be drastically different 
compared with the procedure used in the present work. 
As one could see, the resulting $g_{pp}$ 
are similar (but not identical) and the dependence on the 
s.p. basis will be also reduced (That important feature was
not demonstrated in Ref. \cite{Suh04}, unfortunately). However, as the upper panel 
shows, it is difficult to describe the other component of that decay, 
the  $\beta^+/EC$ amplitude. That is an obvious drawback of the 
QRPA method; it is never meant to describe in detail properties 
of non-collective states. But that is less relevant 
for the description of integral quantities that depend on  
sums over many states. 
 
As pointed out earlier, we always choose the $g_{pp}$ corresponding to
the positive $M^{2\nu}$ value. In the upper panel of Fig.~\ref{0nbbfig:10}
one can clearly see that were we to choose the other possibility, i.e.,
the $g_{pp}$ corresponding to the negative $M^{2\nu}$, the disagreement
with the single beta decay would be considerably worse. 
It is important also to notice that if Pauli principle violation is restored 
(e.g. within the RQRPA) one finds that the solution corresponding to the 
negative value of $M^{2\nu}$ is out of physical interval of $g_{pp}$. 
This has been confirmed also in a schematic model
by presenting the solution of the QRPA with full inclusion
of the Pauli exclusion principle \cite{sim01}.
As pointed out above, it was also found recently \cite{deform}
that for negative value of $M^{2\nu}$ the
correspondence with particle-particle strength from systematic
studies of the single beta decay \cite{homma} is not achieved.

In this section we have summarized the arguments why we believe that
the procedure of adjustment used in the present work is preferable to
the procedure advocated in Ref.\cite{Suh04}. At the same time, we suggest
that there is no fundamental difference between the two; both are used to
fix the fast varying contribution of the $1^+$ multipole. The bulk of the
matrix element is associated with the other multipoles, and their effect is 
much less dependent on relatively small variations of the parameter $g_{pp}$.

As we stressed already above, a substantial part of the differences
in the calculated values of the $0\nu\beta\beta$ decay matrix elements has
its origin not in the choice of $g_{pp}$ or other parameters, but in the neglect
of the short range nucleon-nucleon repulsion and of the induced weak currents
in some papers, and their inclusion in other papers (like the present one).
Until a consensus on the treatment of these physics issues is reached, the
differences in the calculated matrix elements cannot be avoided.

\section{Summary and conclusions}

We have shown that the procedure suggested in our previous work, Ref. \cite{Rod03a}, 
is applicable to essentially all nuclei with known $2\nu\beta\beta$ decay lifetimes .  
Adjusting the strength 
of the particle-particle neutron-proton force $g_{pp}$ in such a way that the 
experimental $2\nu\beta\beta$ decay rate is correctly reproduced removes much 
of the dependence on the size of the single-particle basis and  
whether QRPA or RQRPA is used. Here we also show that the quenching of the axial 
current matrix elements, parameterized by the reduction of the coupling constant 
$g_A$, also leaves the resulting $0\nu\beta\beta$ matrix elements almost unchanged; 
they become insensitive to the variations of parameters describing the short-range 
nucleon-nucleon correlations as well. Thus,  
the resulting $0\nu\beta\beta$ matrix elements acquire well defined values, 
free of essentially arbitrary choices.  
We also present arguments while we believe 
that the chosen procedure of adjusting the interaction is preferable to other 
proposed ways of adjustment. 
 
We then summarize many published QRPA and RQRPA results and discuss their 
similarities and differences. We show that in most, albeit not all, cases these 
differences can be understood. We present an exhaustive list of reasons why 
individual calculated  $0\nu\beta\beta$ nuclear matrix elements, 
evaluated within QRPA or RQRPA, might differ from each other.  
That list then can serve as a guide to readers of the past and future  
papers devoted to the subject. 
Comparison between the results of different QRPA/RQRPA calculations
would be facilitated
if authors of future publications specify in detail what choices of
explicit and implicit adjustable parameters 
they made, and discuss the dependence of their result on their particular 
choice.  We believe that by following these 
suggestions a consensus among the practitioners of QRPA/RQRPA
could be reached and most of the spread between the calculated nuclear matrix elements, 
that causes much confusion in the wider physics community, would be shown 
to be essentially irrelevant. To reach a convergence
of the results obtained using QRPA/RQRPA is clearly just an important step on the way
to reliable and correct $0\nu\beta\beta$ decay nuclear matrix elements. Exploring
the structure of the intermediate odd-odd nuclei by the charge exchange reactions 
would create an opportunity to test the nuclear models more thoroughly. Further
progress in nuclear shell model calculations and in the exploration of the exactly 
solvable models could also point the way towards the ultimate solution
of this important problem.

\acknowledgments 
 
Discussions with Jon Engel are gratefully acknowledged. 
The work of F. \v{S}. and V. R. was supported in part by the Deutsche 
Forschungsgemeinschaft (grants 436 SLK 17/298, TU 7/134-1 and FA67/28-2,
respectively). 
We thank also  the EU ILIAS project under the contract RII3-CT-2004-506222. 
The work of P. V. was  supported by the U.S. DOE, Stanford University, SLAC and KIPAC.

\newpage 

\begin{table}[t] 
  \begin{center} 
    \caption{Averaged $0\nu\beta\beta$ nuclear matrix elements 
 $\langle {M'}^{0\nu} \rangle$ and their variance $\sigma$  (in parentheses) 
evaluated in the RQRPA and QRPA. In column 6 the variance $\varepsilon_{exp.}$  
of the $0\nu\beta\beta$-decay matrix element due to uncertainties in the 
measured $2\nu\beta\beta$-decay half-live $T^{2\nu-exp}_{1/2}$  
is given. $M_{GT}^{exp}$ and $g_A$ 
denote the $2\nu\beta\beta$-decay nuclear matrix element deduced from  
$T^{2\nu-exp}_{1/2}$  and axial-vector coupling constant, respectively.   
In column 7 the  $0\nu\beta\beta$ half-lives 
evaluated with the RQRPA average nuclear matrix element and for assumed 
$\langle m_{\beta\beta} \rangle$ = 50 meV are shown. For $^{136}$Xe there are four 
entries; the upper two use the upper limit of the $2\nu$ matrix element while 
the lower two use the ultimate limit, vanishing $2\nu$ matrix element. 
$^{150}$Nd is included for illustration. It is treated as a spherical nucleus;  
deformation will undoubtedly modify its $0\nu$ matrix element. } 
\label{tab:t12} 
\begin{tabular}{lcccccc} 
\hline\hline 
 Nuclear & $~g_A~~$ & $M_{GT}^{exp}$ & \multicolumn{2}{c} 
{$\langle {M'}^{0\nu} \rangle$} & 
 $~~\varepsilon_{exp.}~~$ &   
$T^{0\nu}_{1/2}$ ($\langle m_{\beta\beta} \rangle$ = 50 meV)  \\ \cline{4-5} 
transition & & $[MeV^{-1}]$ & \hspace{0.3cm} RQRPA \hspace{0.3cm}  
& \hspace{0.3cm} QRPA \hspace{0.3cm}  & & [yrs] \\ 
\hline  
$^{76}Ge\rightarrow {^{76}Se}$ 
          & 1.25 & $0.15\pm 0.006$ & 2.40(0.07) & 2.68(0.06) & $\pm 0.05$   & $2.29_{-0.16}^{+0.17}~10^{27}$ \\ 
          & 1.00 & $0.23\pm 0.01$  & 2.30 (0.04) & 2.48 (0.05) & $\pm 0.05$ & $2.49_{-0.13}^{+0.15}~10^{27}$ \\ 
$^{82}Se\rightarrow {^{82}Kr}$    
          & 1.25 & $0.10\pm 0.009$ & 2.12 (0.10) & 2.36 (0.09) & $\pm 0.04$ & $6.60_{-0.62}^{+0.73}~10^{26}$ \\ 
          & 1.00 & $0.16\pm 0.008$ & 1.91 (0.05) & 2.10 (0.07) & $\pm 0.05$ & $8.13_{-0.57}^{+0.64}~10^{26}$ \\ 
$^{96}Zr\rightarrow {^{96}Mo}$    
          & 1.25 & $0.11^{+0.03}_{-0.06}$ & 0.31 (0.08) & 0.04 (0.10) & $^{+0.20}_{-0.43}$ & $1.45_{-0.96}^\infty~10^{28}$ \\ 
          & 1.00 & $0.17^{+0.05}_{0.06}$ & 0.43 (0.11) & 0.40 (0.02) & $^{+0.25}_{-0.45}$ & $0.77_{-0.48}^\infty~10^{28}$ \\ 
$^{100}Mo\rightarrow {^{100}Ru}$    
           & 1.25 & $0.22\pm 0.01$ & 1.16(0.11) & 1.28(0.09) & $\pm 0.02$ & $1.36_{-0.23}^{+0.30}~10^{27}$ \\ 
           & 1.00 & $0.34\pm 0.015$ & 1.12 (0.09) & 1.24 (0.08) & $\pm 0.02$ & $1.45_{-0.21}^{+0.27}~10^{27}$ \\ 
$^{116}Cd\rightarrow {^{116}Sn}$    
           & 1.25 & $0.12\pm 0.006$ & 1.43 (0.08) & 1.56 (0.10) & $\pm 0.03$ & $8.20_{-0.90}^{+1.07}~10^{26}$ \\ 
           & 1.00 & $0.19\pm 0.009$ & 1.22 (0.07) & 1.31 (0.08) & $\pm 0.02$ & $11.3_{-1.23}^{+1.48}~10^{26}$ \\ 
$^{128}Te\rightarrow {^{128}Xe}$    
           & 1.25 & $0.034\pm 0.012$ & 1.60 (0.11) & 1.73 (0.13) & $\pm 0.09$ & $1.85_{-0.29}^{+0.38}~10^{28}$ \\ 
           & 1.00 & $0.053\pm 0.02$  & 1.37 (0.07) & 1.47 (0.05) & $\pm 0.1$  & $2.52_{-0.40}^{+0.52}~10^{28}$ \\ 
$^{130}Te\rightarrow {^{130}Xe}$    
           & 1.25 & $0.036^{+0.03}_{-0.009}$ & 1.47 (0.15) & 1.55(0.17) & $^{+0.3}_{-0.09}$ & $0.87_{-0.16}^{+0.25}~10^{27}$ \\ 
           & 1.00 & $0.056^{+0.05}_{-0.15}$  & 1.28 (0.08) & 1.36 (0.10) & $^{+0.27}_{-0.08}$ & $1.15_{-0.38}^{+0.23}~10^{27}$ \\ 
$^{136}Xe\rightarrow {^{136}Ba}$    
           & 1.25 & $0.030$ & 0.98(0.09) & 1.03(0.08) & & $1.84_{-0.30}^{+0.39}~10^{27}$ \\ 
           & 1.00 & $0.045$ & 0.90 (0.07) & 0.94 (0.05) & & $2.18_{-0.16}^{+0.25}~10^{27}$ \\ 
           & 1.25 & 0 & 0.73(0.09) & 0.77(0.10) & & $3.32_{-0.69}^{+1.00}~10^{27}$ \\ 
           & 1.00 & 0 & 0.63 (0.07) & 0.67 (0.08) & & $4.45_{-0.85}^{+1.18}~10^{27}$ \\ 
$^{150}Nd\rightarrow {^{150}Sm}$    
           & 1.25  & $0.07^{+0.009}_{-0.03}$ & 2.05 (0.13) & 2.25 (0.16) & $^{+0.07}_{-0.20}$ & $0.92_{-0.12}^{+0.26}~10^{26}$ \\ 
           & 1.00  & $0.11^{+0.014}_{-0.05}$ & 1.79 (0.05) & 1.96 (0.04) & $^{+0.06}_{-0.19}$ & $1.21_{-0.10}^{+0.32}~10^{26}$ \\ 
\hline\hline 
\end{tabular} 
  \end{center} 
\end{table} 
 
\begin{table}[t]
  \begin{center}
    \caption{Neutrinoless double beta decay matrix element ${M'}^{0\nu}$
calculated within QRPA and RQRPA approaches. The way of fixing the
particle-particle interaction strength is indicated: a given value,
to $2\nu\beta\beta$-decay half-life, to $\beta$ decays.
The nuclear radius is $R = r_0 A^{1/3}$.
Notation:
EVZ-88 = Engel, Vogel, Zirnbauer\protect\cite{Engel88} 
(for $\alpha_1'$ = -390),
MBK-89 = Muto, Bender, Klapdor\protect\cite{mut89},
T-91 = Tomoda \protect\cite{tom91},
SKF-91 = Suhonen, Khadkikhar, Faessler\protect\cite{suh91},
PSVF-96 = Pantis, \v Simkovic, Vergados, Faessler\protect\cite{pan96},
AS-98 = Aunola, Suhonen\protect\cite{au98}, results for Woods-Saxon (WS)
and adjusted Woods-Saxon (AWS) bases,
SPVF-99 = \v Simkovic, Pantis, Vergados, Faessler\protect\cite{si99},
SK-01 = Stoica, Klapdor-Kleingrothaus\protect\cite{SK01}
(the second line by SK-01 results corresponds to slightly larger
model space),
CS-03 = Civitarese, Suhonen\protect\cite{CS03}.
}
\label{tab:comp}
\renewcommand{\arraystretch}{1.2}
\begin{tabular}{lcccc|cccccccccc}
\hline\hline
Ref.  & Method & $r_0$ & $g_{pp}$ &  $g_A$ & \multicolumn{9}{c}
{ ${M'}^{0\nu}$}\\\cline{6-14}
 & & [fm] & & &  $^{76}Ge$ & $^{82}Se$ & $^{96}Zr$ & $^{100}Mo$
 & $^{116}Cd$ & $^{128}Te$ & $^{130}Te$ & $^{136}Xe$ & $^{150}Nd$ \\ \hline
 & & & & & \multicolumn{9}{c}{Without higher order terms of nucleon current}
\\
 & & & & & \multicolumn{9}{c}{Differences understandable} \\
EVZ-88  & QRPA  & 1.1 & $\beta$  & 1.0
  & 1.3 & 1.0 & 0.8  & 1.8 &       & 2.4 & 2.2 & 1.0 &  \\
MBK-89  & QRPA  & 1.1 & $\beta$  & 1.25
  & 3.84 & 3.59 &      & 1.94 &      & 3.93 & 3.18 & 1.45 & 5.57 \\
T-91   & QRPA  & 1.1 & 1.0  & 1.0
  & 2.86 & 2.59 &      & 3.14 &      & 2.55 & 2.22 & 1.27 & 3.47 \\
       &    &     &     & 1.25
  & 3.97 & 3.60 &      & 4.30 &      & 3.53 & 3.07 & 1.74 & 4.80 \\
SKF-91   & QRPA  &  1.1 & 1.0 & 1.00
  & 3.37 & 2.72 &      &      &      & 3.17 & 2.94 & 1.71  &      \\
         &   &      &      & 1.25
  & 4.55 & 3.71 &      &      &      & 4.24 & 3.95 & 2.31  &      \\
PSVF-96 & QRPA  & 1.1 & 1.0  & 1.25
  & 3.04 & 2.23 & 2.41 & 1.09 & 0.94 & 2.48 & 2.33 & 1.55 &      \\
SPVF-99 & RQRPA & 1.1 & 1.0  & 1.00
  & 4.05 & 3.82 & 2.24 & 4.58 & 2.86 & 3.38 & 2.87 & 1.20 & 5.15 \\
        &    &     &      & 1.25
  & 3.60 & 3.40 & 1.99 & 4.12 & 2.58 & 2.96 & 2.50 & 1.02 & 4.51 \\
present & QRPA  & 1.1 & $2\nu\beta\beta$ & 1.25
  & 3.35 & 2.95 &      & 1.83 &      & 2.32 & 1.98 & 1.30 & 3.10 \\
 & & & & & \multicolumn{9}{c}{Results discussed} \\
SK-01    & QRPA  &  ?  & $2\nu\beta\beta$ & ?
  & 4.45 & 5.60 &  4.16 & 5.37 & 3.99 & 4.84 & 4.73 & 1.69 &     \\
         &   &      &      &
  & 1.71 & 4.71 &  2.75 & 3.81 & 2.85 & 3.43 & 3.77 & 1.35 &     \\
         & RQRPA  &  ?   & $2\nu\beta\beta$ & ?
  & 3.74 & 4.30 &  3.01 & 4.36 & 3.61 & 4.29 & 4.55 & 1.57 &     \\
         &   &      &      &
  & 1.87 & 2.70 &  2.72 & 3.40 & 3.39 & 2.83 & 3.00 & 1.02 &     \\
$AS_{WS}$-98 & QRPA  & 1.1 & $\beta$ & 1.00
  & 3.98 & 3.69 &  2.88  &     & 2.21  &  & 4.62 & 2.49 &  \\
        &         &     &             & 1.25
  & 5.30 & 4.93 &  3.85  &     & 2.93  &  & 6.15 & 3.34 &  \\
$AS_{AWS}$-98 & QRPA  & 1.1 & $\beta$ & 1.00
  & 4.85 & 3.61 &  3.70  &     & 3.97  &  & 3.81 & 2.15 &  \\
        &         &     &             & 1.25
  & 6.44 & 4.82 &  3.96  &     & 5.25  &  & 5.05 & 2.84 &  \\
CS-03   & QRPA  & ?  &  $\beta$ & 1.25
  & 3.33 & 3.44 & 3.55 & 2.97 & 3.75 &      & 3.49 & 4.64 &      \\
 & & & & & \multicolumn{9}{c}{With higher order terms of nucleon current} \\
SPVF-99 & RQRPA & 1.1 & 1.0  & 1.25
  & 2.80 & 2.64 & 1.49 & 3.21 & 2.05 & 2.17 & 1.80 & 0.66 & 3.33  \\
present & QRPA  & 1.1 & $2\nu\beta\beta$ & 1.00
  & 2.48 & 2.10 & 0.40 & 1.24 & 1.31 & 1.47 & 1.36 & 0.94 & 1.96  \\
         &      &     &               & 1.25
  & 2.68 & 2.36 & 0.04 & 1.28 & 1.56 & 1.73 & 1.55 & 1.03 & 2.25  \\
         & RQRPA & 1.1 & $2\nu\beta\beta$ & 1.00
  & 2.30 & 1.91 & 0.43 & 1.12 & 1.22 & 1.37 & 1.28 & 0.90 & 1.79  \\
        &       &    &                 & 1.25
  & 2.40 & 2.12 & 0.31 & 1.16 & 1.43 & 1.60 & 1.47 & 0.98 & 2.05  \\
\hline\hline
\end{tabular}
  \end{center}
\end{table}

\newpage 
 
 
\begin{thebibliography}{99}
\bibitem{foot1} $0\nu\beta\beta$ decay can 
be also mediated by the exchange of various heavy particles. For a review 
of this possibility see \cite{FS98}. Relation to the lepton flavor violation 
has been suggested in \cite{cir04}.
\bibitem{foot2} Similar conclusions were reached by J. Engel earlier in a study 
involving an exactly solvable model. We appreciate his input.
\end{thebibliography}
\end{document}